\documentclass[reprint,showpacs,preprintnumbers,amsmath,amssymb,prl]{revtex4-1}
\usepackage{amsmath}
\usepackage{amsfonts}
\usepackage{amsthm}
\usepackage{dsfont}
\usepackage{graphics}
\usepackage{graphicx,bm}
\usepackage[caption=false]{subfig}
\usepackage{amssymb}
\usepackage{mathrsfs}
\usepackage{braket}
\usepackage{float}
\usepackage{color}
\usepackage{url}
\usepackage[abs]{overpic}
\usepackage[usenames,dvipsnames]{xcolor}
\usepackage{commath}
\usepackage{subfig}
\usepackage{multirow}
\usepackage{verbatim}
\usepackage{array}
\usepackage[hidelinks]{hyperref}

\begin{document}

\title{Dirac quantum walks on triangular and honeycomb lattices}

\author{Gareth Jay}
\email{gareth.jay@uwa.edu.au}
\affiliation{Physics Department, The University of Western Australia, Perth, WA 6009, Australia}
\author{Fabrice Debbasch}
\email{fabrice.debbasch@gmail.com}
\affiliation{Sorbonne Universit\'e, Observatoire de Paris, Universit\'e PSL, CNRS, LERMA, F-75005, {\sl Paris}, France}
\author{Jingbo B. Wang}
\email{jingbo.wang@uwa.edu.au}
\affiliation{Physics Department, The University of Western Australia, Perth, WA6009, Australia}

\date{\today}
\begin{abstract}
In this paper, we present a detailed study on discrete-time Dirac quantum walks (DQWs) on triangular and honeycomb lattices. At the continuous limit, these DQWs coincide with the Dirac equation. Their differences in the discrete regime are analyzed through the dispersion relations, with special emphasis on Zitterbewegung. An extension which couples these walks to arbitrary discrete electromagnetic field is also proposed and the resulting Bloch oscillations are discussed.

\end{abstract}
%\pacs{03.67.-a, 47.37.+q, 47.40.-x, 67.10.-j}
\maketitle
\section{I. Introduction}

Quantum walks were first considered by Feynman in studying possible discretizations for the Dirac path integral \cite{feynman2010quantum,schweber1986feynman}. They were later introduced as quantum automata in a systematic way by Aharonov et al. \cite{aharonov1993quantum} and Myers et al. \cite{meyer1996quantum}. Quantum walks are simple models of coherent quantum transport on discrete structures such as graphs and lattices. They have attracted a lot of attention in quantum information and algorithmic development \cite{ambainis2007quantum,magniez2011search,ManouchehriWang2014}. They can also be used as quantum simulators \cite{Strauch2006, Strauch2007, Kurzynski2008, Chandrashekar2013, Shikano2013, Arrighi2014,arrighi2016quantum, Molfetta2014, perez2016asymptotic}, with the lattice now representing a discretization of continuous space. On the other hand, in a more adventurous way, quantum walks may represent a potentially realistic discrete space underlying the apparently continuous physical universe \cite{bisio2016special}.

It has been shown recently that several Discrete-Time Quantum Walks defined on regular square lattices simulate the Dirac dynamics in various space-time dimensions and that these walks (termed as DQWs) can be coupled to various discrete gauge fields \cite{di2013quantum,di2014quantum,arnault2016landau,arnault2016quantum,arnault2016quantum2,arnault2017quantum,bisio2015quantum}. In particular, some DQWs can be coupled to discrete electric and magnetic fields in a gauge invariant manner \cite{di2014quantum,arnault2016landau,arnault2016quantum}, and magnetic confinement as well as Bloch oscillations have been observed. A natural, yet unanswered question is: Can similar results be replicated on non-square lattices and, if so, which aspects depend on the choice of lattice, and which do not? The present paper is a first step in answering this question.

We focus on the transport properties of quantum walks in $(1 + 2)D$ space-time dimensions and propose four new unitary DQWs defined on triangular and honeycomb lattices. Two of these DQWs are defined on the same lattice made out of equilateral triangles, the third DQW is defined on a lattice of isosceles triangles, and the last is on a  hexagonal honeycomb lattice. At the continuous limit, all four walks are identical and coincide with the free Dirac equation. However, the walks greatly differ outside this limit and  the differences are analyzed through the dispersion relations~\cite{Valcarcel2010, Ahlbrecht2011, Hinarejos2013}, with special emphasis on Zitterbewegung~\cite{Schliemann2005, Gerritsma2010, Bisio2013}. We also discuss how these walks can be extended to include a gauge-invariant coupling to arbitrary electromagnetic fields. The extension is built in full for the simplest walk defined on the equilateral triangle lattice, for which Bloch oscillations~\cite{Bloch1929, Tamascelli2016} are also addressed. These results show that DQWs defined on
more general lattices than the square lattice can be used to study the free Dirac dynamics and that coupling the walks to arbitrary electromagnetic fields in a gauge invariant manner is also possible.

%Finally, the Appendix introduces a fourth DTQW whose construction is inspired  by one of the three DTDQWs presented in the main part of this article. This fourth walk couples close neighbours on the honeycomb lattice and can therefore  be used as a model of transport in graphene.

\section{II. Four free quantum walks on non-square lattices}

\subsection{1. Six step walk on the equilateral triangular lattice}

An arbitrary element $U$ of $U(2)$ can be parametrized by four angles:
\begin{equation}
U(\alpha, \xi, \zeta, \theta) = e^{i \alpha}
\begin{pmatrix}
e^{i \xi} \cos \theta & e^{i \zeta} \sin \theta \\
- e^{- i \zeta} \sin \theta & e^{- i \xi} \cos \theta
\end{pmatrix}.
\end{equation}
Let $(x,y)$ be orthonormal coordinates on the plane and focus first on the regular lattice made of equilateral triangles of side $\epsilon$. Choose one site $S$ at position ${\bf X} = (x,y)$ and let its six neighbours $N_1({\bf X})$, $N_2({\bf X})$, $N_3({\bf X})$, $N_4({\bf X})$, $N_5({\bf X})$ and $N_6({\bf X})$ have the respective coordinates ${\bf X} _1 = (x + \epsilon, y)$, ${\bf X} _2 = (x + \epsilon/2, y + \sqrt{3}\epsilon/2)$, ${\bf X} _3 = (x - \epsilon/2, y + \sqrt{3}\epsilon/2)$, ${\bf X} _4 = (x -\epsilon , y)$,
${\bf X} _5 = (x - \epsilon/2, y - \sqrt{3}\epsilon/2)$, ${\bf X} _6 = (x + \epsilon/2, y - \sqrt{3}\epsilon/2)$. The six corresponding translation operators $S_j$ are defined by
$(S_j \Psi )({\bf X} ) = (\psi^L (N_j({\bf X} )),\psi^R (N_{j+1}({\bf X} )))^\top$, where $N_7=N_1$. Consider the DQW defined by
$\Psi(t + \Delta t) = W_0 \Psi (t)$ with $W_0 = \Pi_{j = 1}^6 W_j$ where
\begin{eqnarray}
W_j & = & R_j^{-1} U_j S_j R_j, \nonumber \\
R_j & = & U(0, \pi/2, 0, \pi/12 + (j - 1) \pi/6), \nonumber \\
R_j^{-1} & = & U(0, - \pi/2, 0, - \pi/12 - (j - 1) \pi/6), \nonumber \\
U_j  & = & U(0, 0, 0, 0) = 1.
\end{eqnarray}
Introducing the operators $U_j$ seems useless because, at this stage, they all coincide with the unit operator. This is so because, as explained in Section IV below, these operators actually code for an electromagnetic field acting on the walk and we are considering only free walks in the present section. The more general case is described in Section IV.

Set now $\Delta t = 3\epsilon/2$ and let $\epsilon$ tend to zero. The formal limit of this DQW exists and coincides then with the mass-less Dirac equation $\gamma^\mu \partial_{\mu} \Psi = 0$ with $x^0 = t$, $x^1 = x$, $x^2 = y$ and
\begin{eqnarray}
\gamma^0 & = & \sigma_1 =
\begin{pmatrix}
0 & 1 \\
1 & 0
\end{pmatrix}, \nonumber \\
\gamma^1 & = & -i \sigma_3 =
\begin{pmatrix}
-i & 0 \\
0 & i
\end{pmatrix}, \nonumber \\
\gamma^2 & = & -i \sigma_2 =
\begin{pmatrix}
0& -1 \\
1 & 0
\end{pmatrix}.
\end{eqnarray}
A non-vanishing mass $m$ can be added by replacing $W_0$ by $W_m = U_m W_0$ with $U_m = U(0, 0, -\pi/2, 3\epsilon m/2)$.

\subsection{2. Three step walk on the equilateral triangular lattice}

The walk just presented approximates the Dirac equation in six steps. It is possible to approximate the Dirac equation by a three-step walk defined on the same equilateral lattice. Consider indeed the DQW defined on the same regular triangular lattice by $\Psi(t+\Delta t)={\tilde W}_0 \Psi(t)$ where
\begin{eqnarray}
  {\tilde W}_0 &=& {\tilde R}_2^{-1} {U}_3 {\tilde S}_3 {\tilde R}_2 {\tilde R}_1^{-1} { U}_2 {\tilde S}_2 {\tilde R}_1 {U}_1 {\tilde S}_1, \nonumber \\
  {\tilde R}_j &=& U(0,\pi/2,0,j\pi/6), \nonumber \\
  {\tilde R}_j^{-1} &=& U(0,-\pi/2,0,-j\pi/6), \nonumber \\
  ({\tilde S}_j\Psi)({\bf X}) &=& (\psi^L (N_j({\bf X} )),\psi^R (N_{j+3}({\bf X} )))^\top,
\end{eqnarray}
and ${U}_j=U(0,0,0,0)$ as before. This walk with $\Delta t = 3\epsilon/2$, and letting $\epsilon$ tend to zero has a formal limit that coincides with the same mass-less Dirac equation as before, except with the gamma matrices this time being
\begin{eqnarray}
\gamma^0 & = & \sigma_1 =
\begin{pmatrix}
0 & 1 \\
1 & 0
\end{pmatrix}, \nonumber \\
\gamma^1 & = & i \sigma_2 =
\begin{pmatrix}
0 & 1 \\
-1 & 0
\end{pmatrix}, \nonumber \\
\gamma^2 & = & -i \sigma_3 =
\begin{pmatrix}
-i & 0 \\
0 & i
\end{pmatrix}.
\end{eqnarray}
A mass $m$ can be added the same way as the other triangular lattice walk above.

\subsection{3. Three step walk on the triangular isosceles lattice}

We introduce a closely related walk, which approximates the Dirac equation also in three steps, but is defined on an isosceles triangular lattice. Consider a lattice of isosceles triangles of base $\epsilon$ and perpendicular height $\epsilon/2$. Each point ${\bf X} = (x,y)$ now has six neighbours defined with the respective coordinates ${\bf X} _1 = (x + \epsilon, y)$, ${\bf X} _2 = (x + \epsilon/2, y + \epsilon/2)$, ${\bf X} _3 = (x - \epsilon/2, y + \epsilon/2)$, ${\bf X} _4 = (x -\epsilon , y)$,
${\bf X} _5 = (x - \epsilon/2, y - \epsilon/2)$, ${\bf X} _6 = (x + \epsilon/2, y - \epsilon/2)$. The corresponding translation operators ${\hat S}_i$ are defined similar to the 3-step walk above. That is, $({\hat S}_j\Psi)({\bf X})=(\psi^L (N_j({\bf X} )),\psi^R (N_{j+3}({\bf X} )))^\top$. The DQW is then defined as $\Psi(t+\Delta t)= {\hat W}_0 \Psi (t)$ where ${\hat W}_0={\hat W}_3{\hat W}_2{\hat W}_1$ with
\begin{eqnarray}
{\hat W}_j & = & {\hat U}_j {\hat S}_j, \nonumber \\
{\hat U}_j & = & U(0, 0, -\pi/2, (j-2)\pi/4).
\end{eqnarray}
With the choice $\Delta t=\epsilon$, this DQW tends to the Dirac equation with the gamma matrices now being $\gamma^0=\sigma_1$, $\gamma^1=i\sigma_2$ and $\gamma^2=i\sigma_3$. To add a non-vanishing mass $m$, one can for example replace ${\hat U}_1$  with ${\hat U}_m=U(0,0,-\pi/2,-\pi/4+m\epsilon)$.

\begin{figure*}[t]
\begin{tabular}{cccc}
  \subfloat[$m=0$]{%
  \includegraphics[width=.3\linewidth]{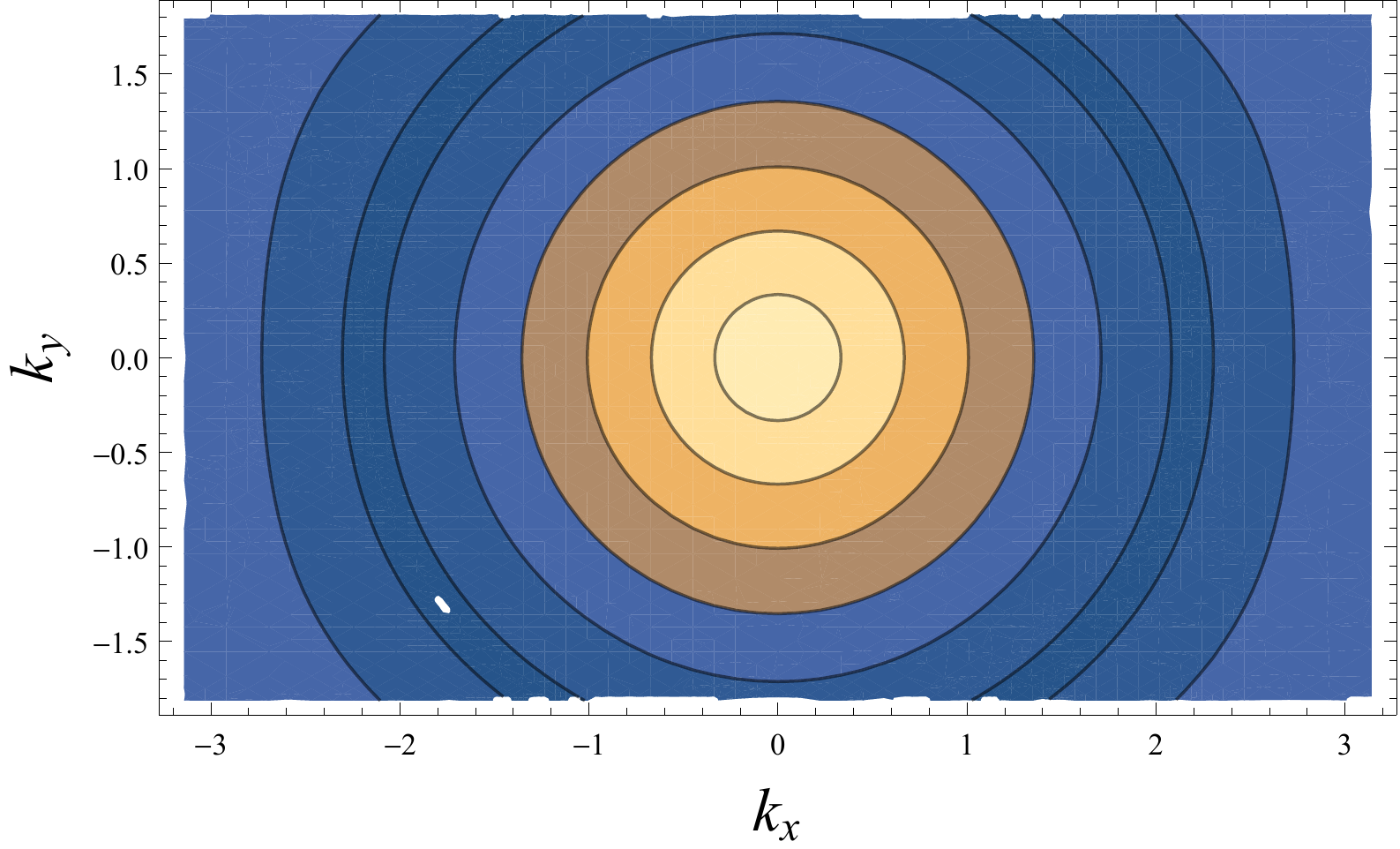}%
}&
\subfloat[$m=\pi/3$]{%
  \includegraphics[width=.3\linewidth]{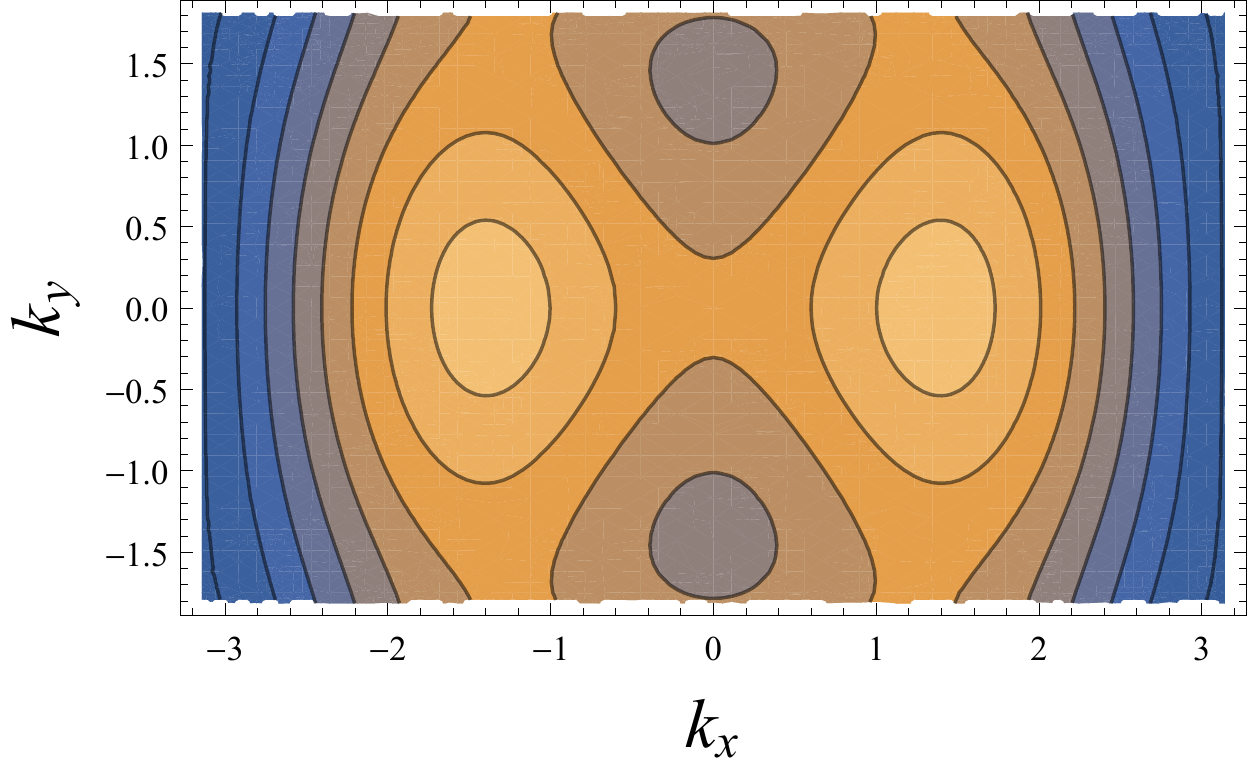}%
}&
\subfloat[$m=\pi/2$]{%
  \includegraphics[width=.3\linewidth]{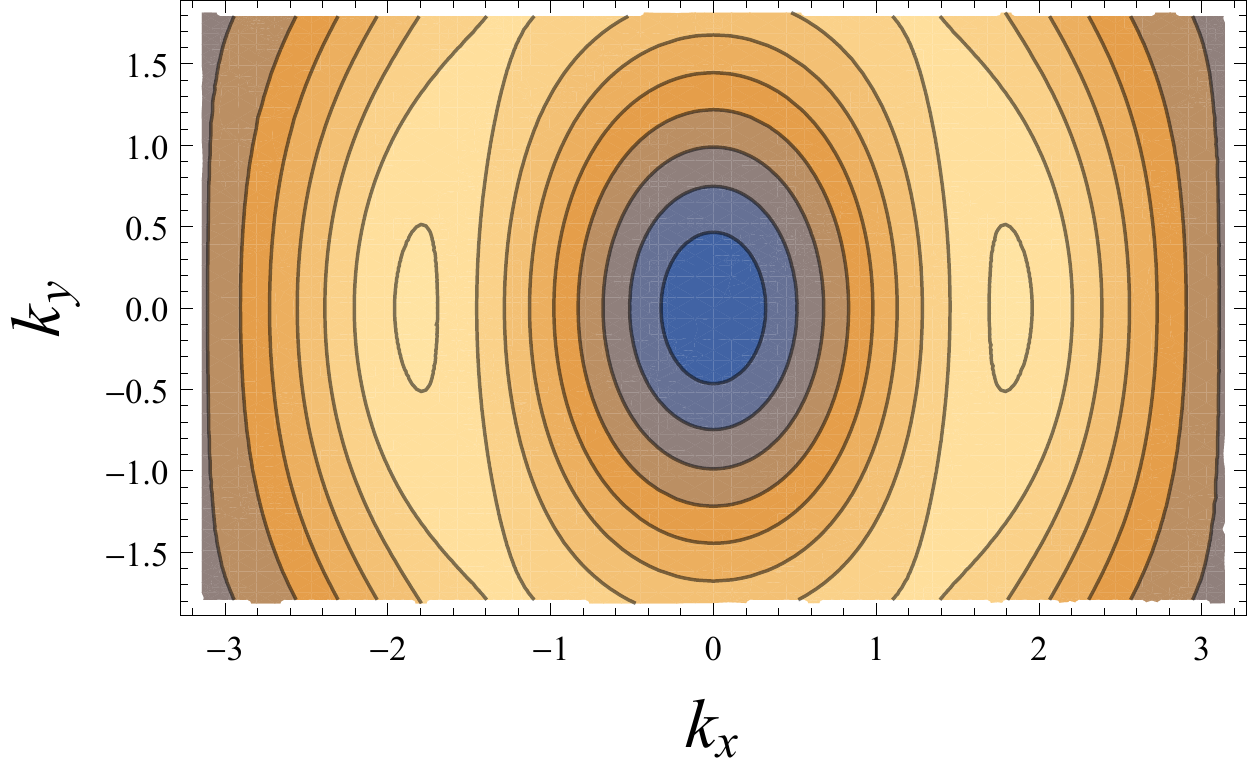}%
}&
\multirow{-3}[2.5]{*}{\subfloat{
  \includegraphics[height=5cm]{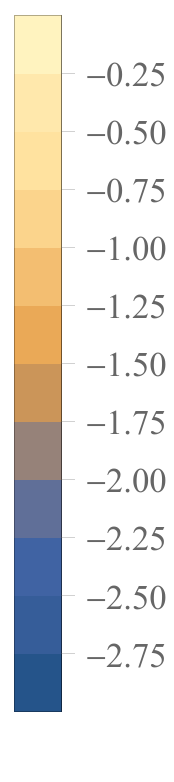}}}\\
\subfloat[$m=2\pi/3$]{%
  \includegraphics[width=.3\linewidth]{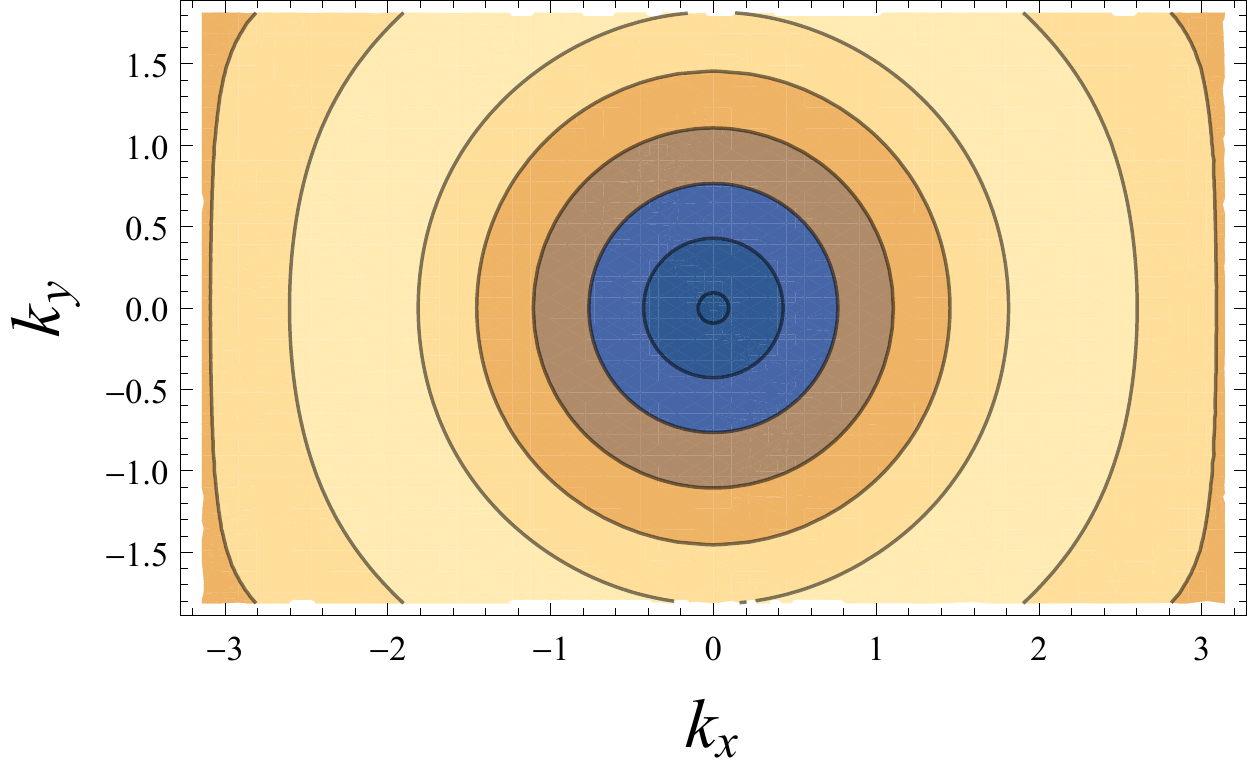}%
}&
\subfloat[$m=\pi$]{%
  \includegraphics[width=.3\linewidth]{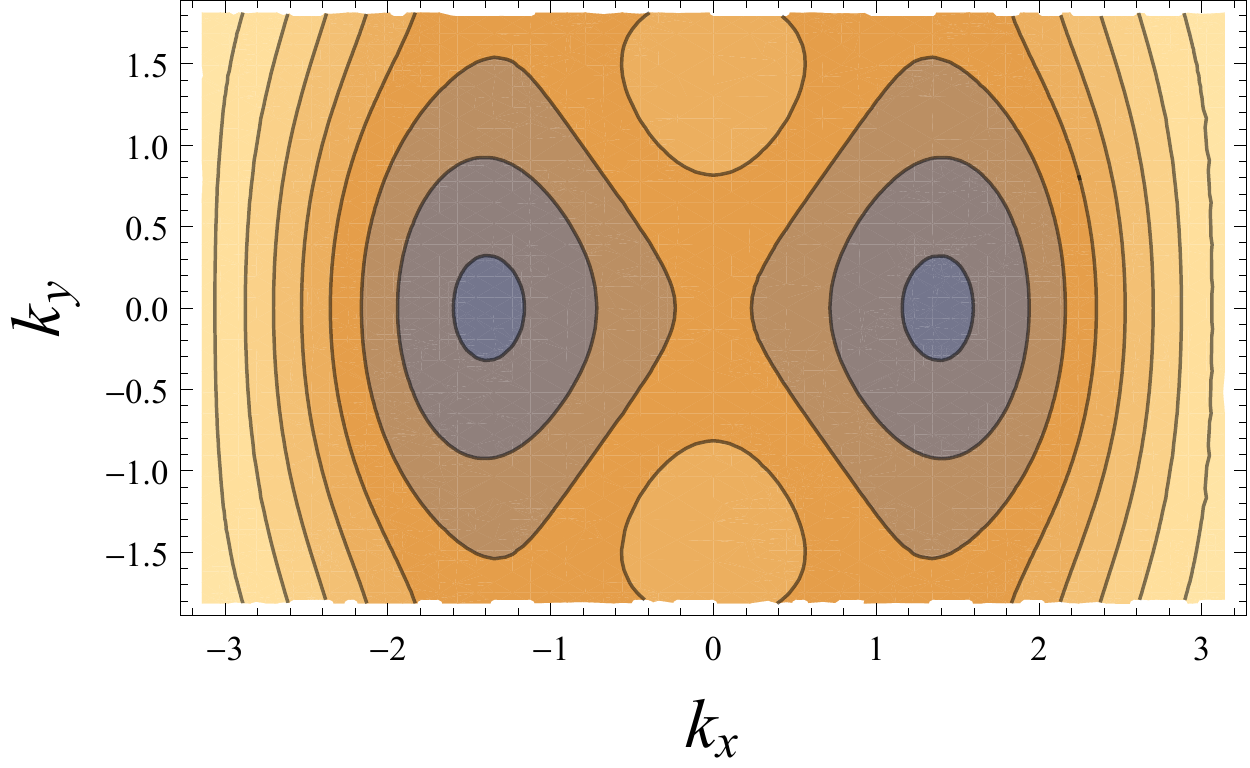}%
}&
\subfloat[$m=4\pi/3$]{%
  \includegraphics[width=.3\linewidth]{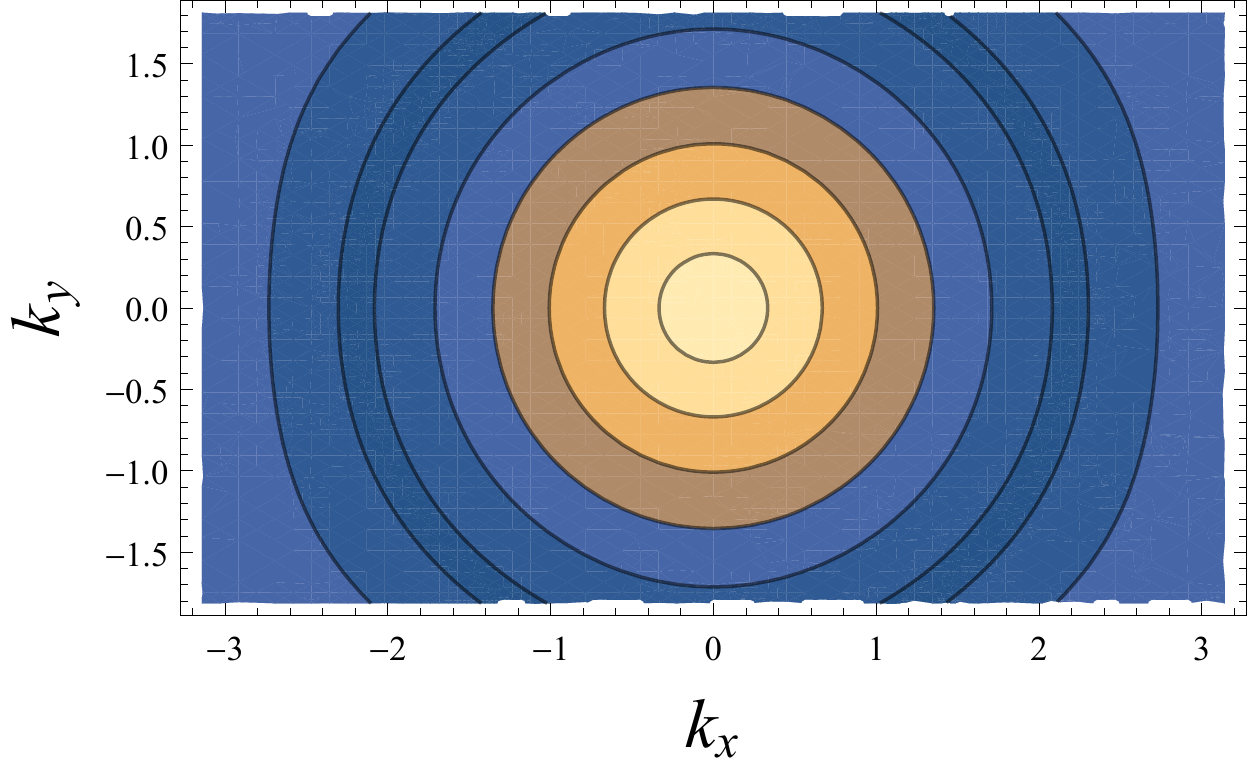}
}\\
\end{tabular}

\caption{Dispersion Relations $\omega(\bf{k})$ for six-step DQW on the equilateral triangular lattice}
\label{6StepEqTri}
\end{figure*}
\begin{figure*}[t]
\begin{tabular}{cccc}
  \subfloat[$m=0$]{%
  \includegraphics[width=.3\linewidth]{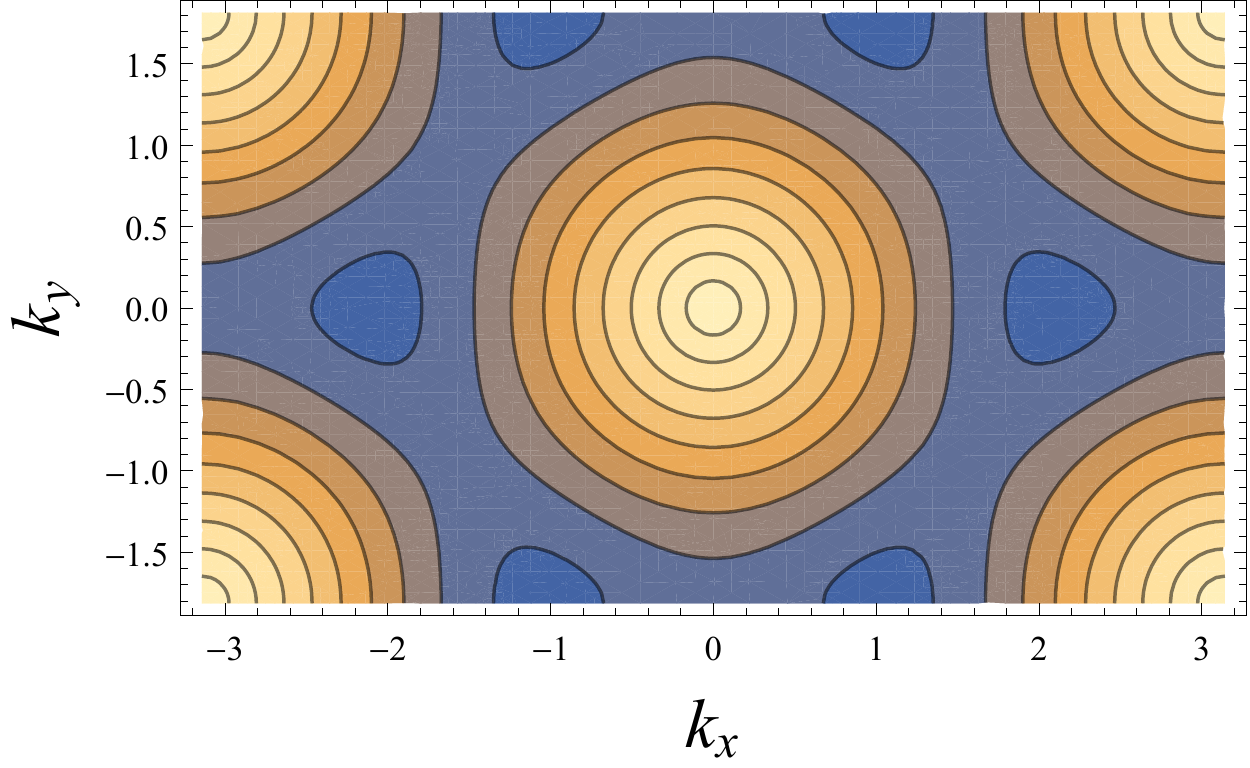}%
}&
\subfloat[$m=\pi/3$]{%
  \includegraphics[width=.3\linewidth]{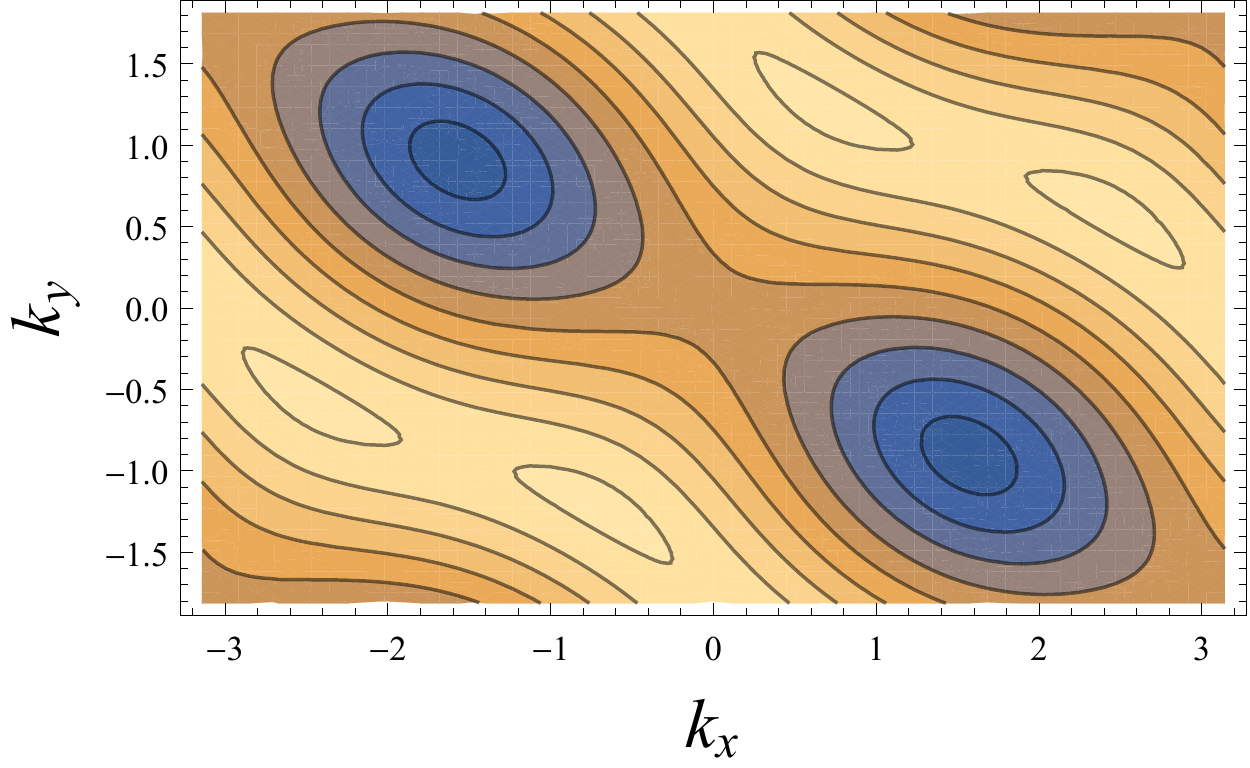}%
}&
\subfloat[$m=\pi/2$]{%
  \includegraphics[width=.3\linewidth]{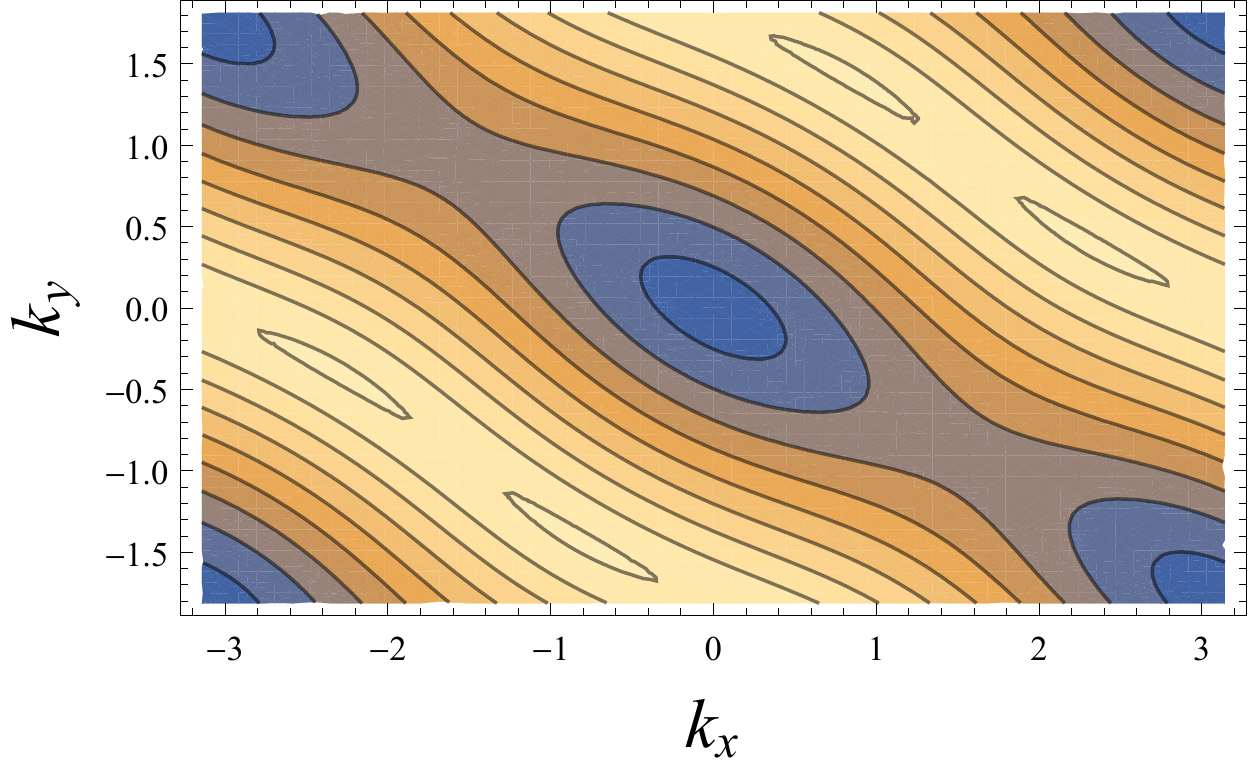}%
}&
\multirow{-3}[2.5]{*}{\subfloat{
  \includegraphics[height=5cm]{Legend}}}\\
\subfloat[$m=2\pi/3$]{%
  \includegraphics[width=.3\linewidth]{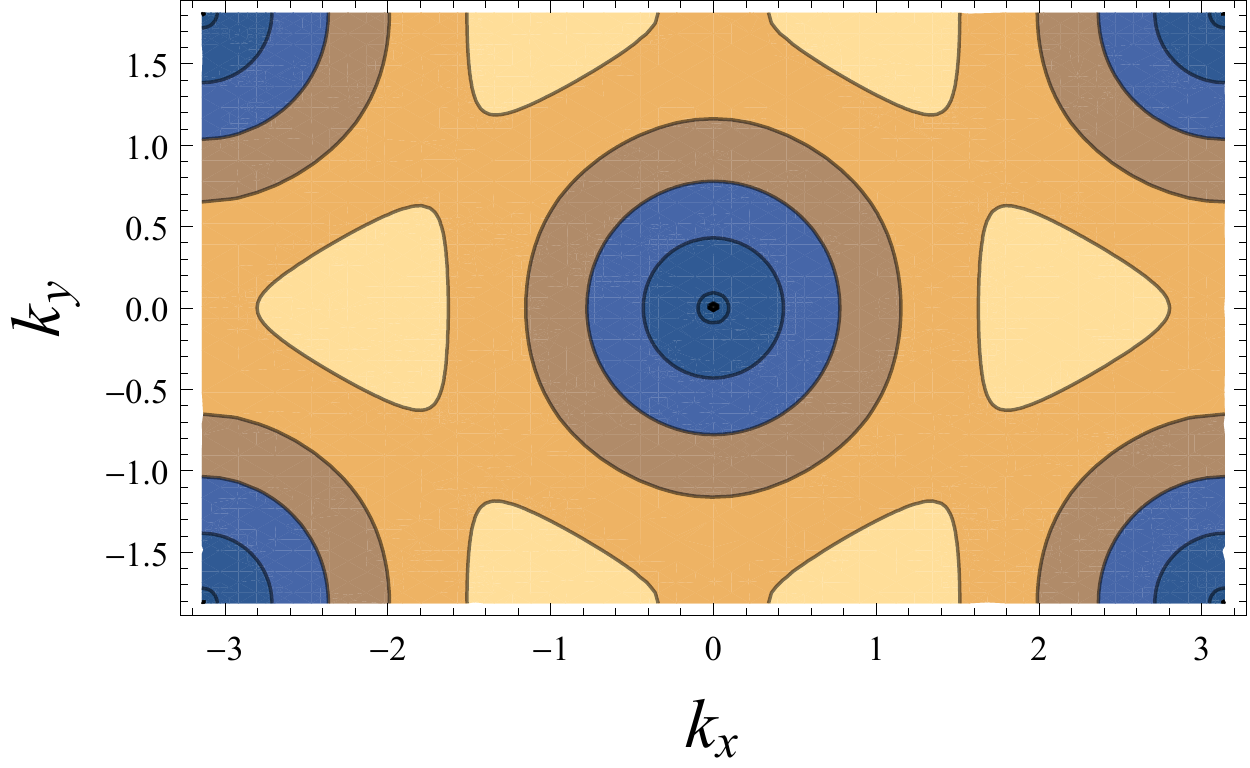}%
}&
\subfloat[$m=\pi$]{%
  \includegraphics[width=.3\linewidth]{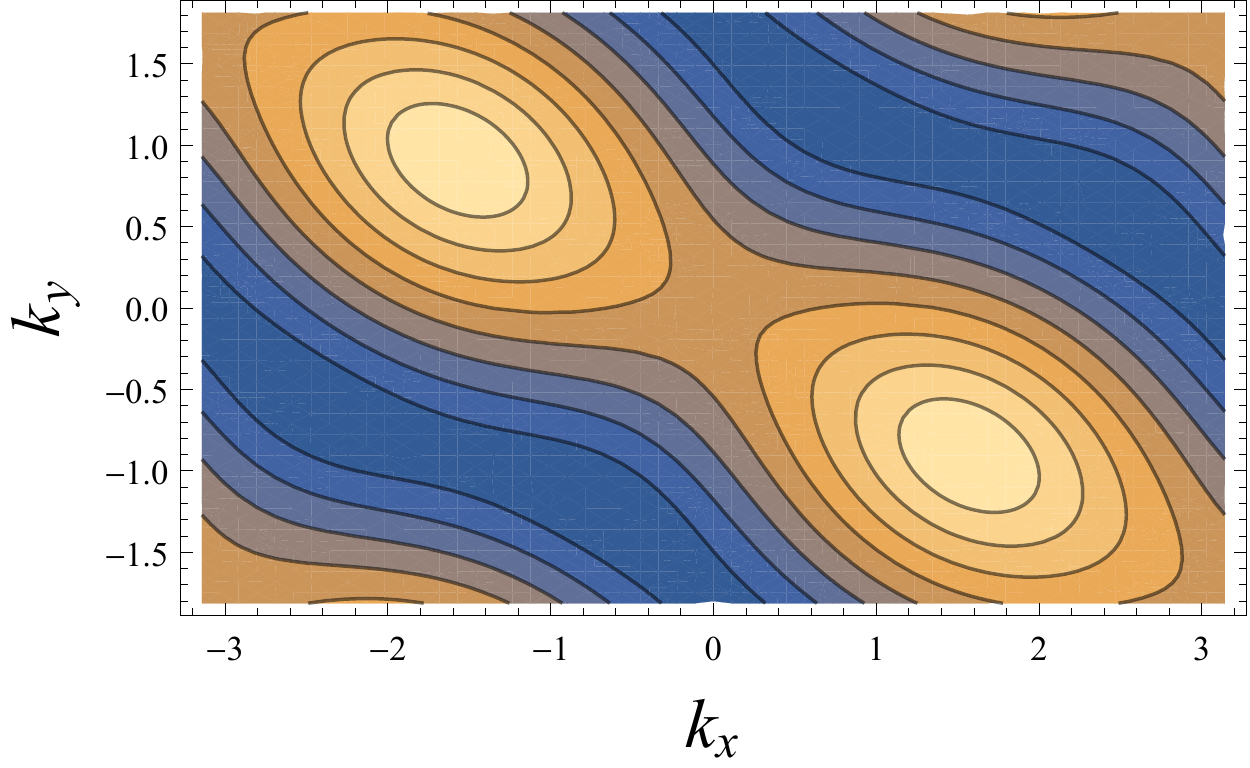}%
}&
\subfloat[$m=4\pi/3$]{%
  \includegraphics[width=.3\linewidth]{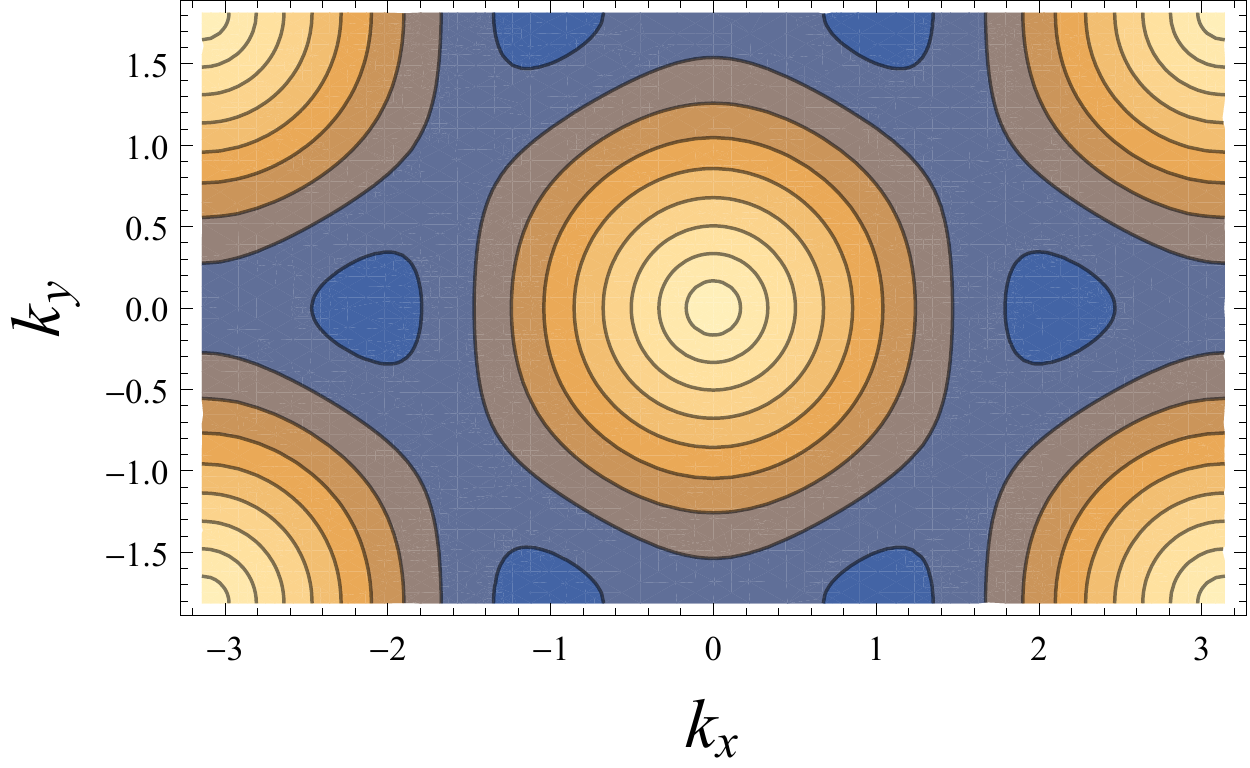}
}\\
\end{tabular}\caption{Dispersion Relations $\omega(\bf{k})$ for three-step DQW on the equilateral triangular lattice}
\label{3StepEqTri}
\end{figure*}

\subsection{4. Three step walk on the hexagonal honeycomb lattice}

Let $(x, y)$ be orthonormal coordinates on the plane and consider a regular hexagonal honeycomb lattice of sides $\epsilon$.
%Each site has three neighbors.
Choose one site $S$ at position ${\bf X} = (x, y)$ and focus on the three neighbours $N_1 ({\bf X} )$, $N_2({\bf X} )$ and $N_3({\bf X} )$ with respective coordinates ${\bf X} _1 = ( x + \epsilon, y)$, ${\bf X} _2 = (x - \epsilon/2, y + \sqrt{3}\epsilon/2)$, ${\bf X} _3 = (x - \epsilon/2, y - \sqrt{3}\epsilon/2)$. Let $\Psi=(\psi^L,\psi^R)^\top$ be a two-component spinor defined on the lattice and define three translation operators by
\begin{eqnarray}
  (S_1 \Psi) ({\bf X} ) &=& (\psi^L (N_1({\bf X} )),\psi^R (N_2({\bf X} )))^\top, \nonumber \\
  (S_2 \Psi) ({\bf X} ) &=& (\psi^L (N_2({\bf X} )),\psi^R (N_3({\bf X} )))^\top, \nonumber \\
  (S_3 \Psi) ({\bf X} ) &=& (\psi^L (N_3({\bf X} )),\psi^R (N_1({\bf X} )))^\top.
\end{eqnarray}
Consider now the DQW defined by $\Psi(t + \Delta t) = W_0 \Psi (t)$ with $W_0 = \Pi_{j = 1}^3 W_j$ where
\begin{eqnarray}
W_j & = & U_j^{-1} S_j U_j, \nonumber \\
U_j & = & U(0, \pi/2, 0, \pi/6 + (j - 1) \pi/3), \nonumber \\
U_j^{-1} & = & U(0, - \pi/2, 0, - \pi/6 - (j - 1) \pi/3).
\end{eqnarray}
This DQW couples sites which are close neighbours on a honeycomb lattice.

Set now $\Delta t = 3\sqrt{3}\epsilon/4$ and let $\epsilon$ tend to zero. The formal limit of this DQW exists and coincides then with the mass-less Dirac equation $\gamma^\mu \partial_{\mu} \Psi = 0$ with $x^0 = t$, $x^1 = x$, $x^2 = y$ and
\begin{eqnarray}
\gamma^0 & = & \sigma_1 =
\begin{pmatrix}
0 & 1 \\
1 & 0
\end{pmatrix}, \nonumber \\
\gamma^1 & = & -i \sigma_3 =
\begin{pmatrix}
-i & 0 \\
0 & i
\end{pmatrix}, \nonumber \\
\gamma^2 & = & -i \sigma_2 =
\begin{pmatrix}
0& -1 \\
1 & 0
\end{pmatrix}.
\end{eqnarray}
A mass $m$ can be added to the DQW by replacing $W_0$ by $W_m = U_m W_0$ with $U_m = U(0, 0, -\pi/2, 3\sqrt{3}\epsilon m/4)$.

\begin{figure*}[t]
\begin{tabular}{cccc}
  \subfloat[$m=0$]{%
  \includegraphics[width=.3\linewidth]{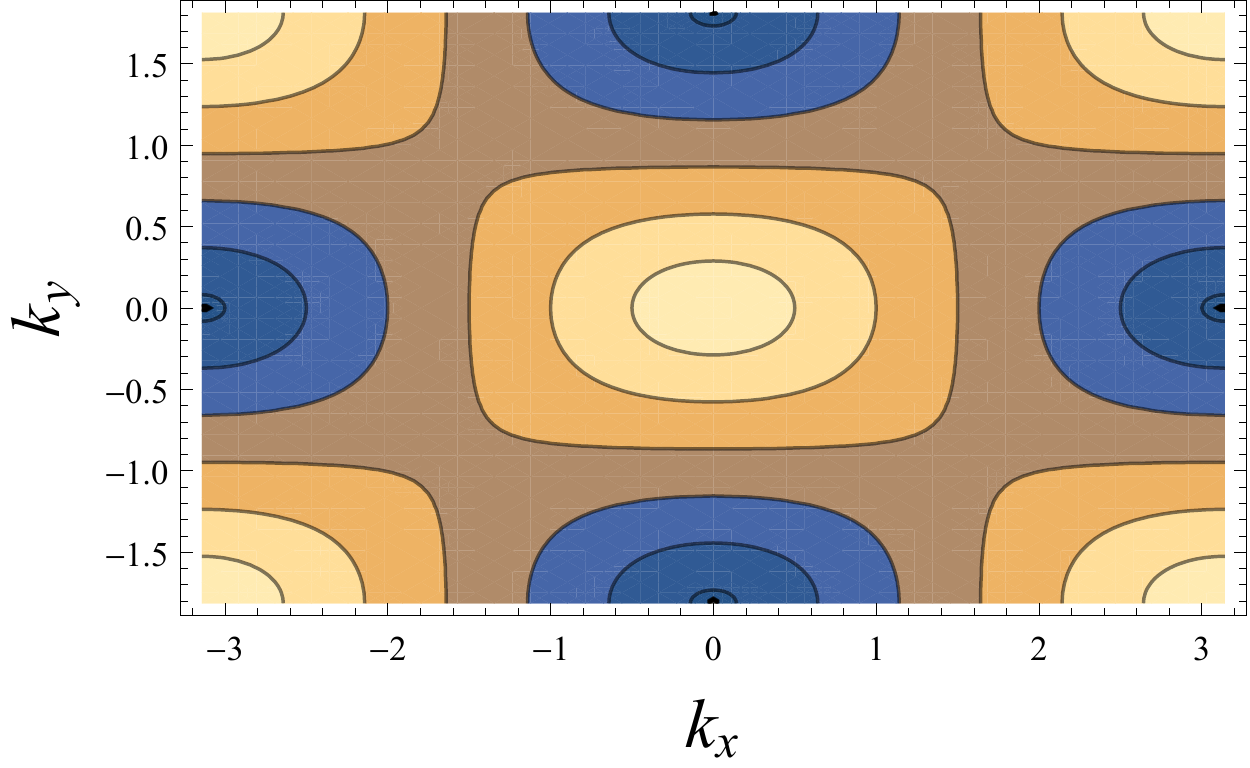}%
}&
\subfloat[$m=\pi/3$]{%
  \includegraphics[width=.3\linewidth]{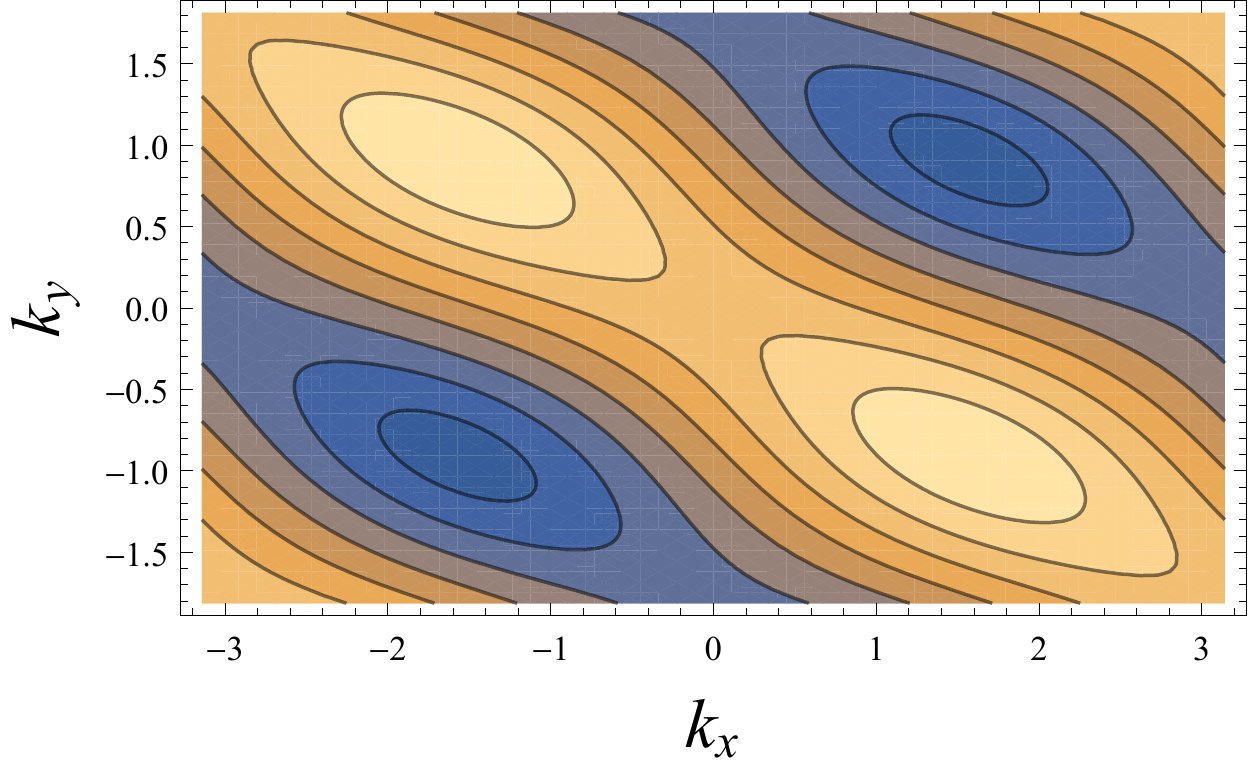}%
}&
\subfloat[$m=\pi/2$]{%
  \includegraphics[width=.3\linewidth]{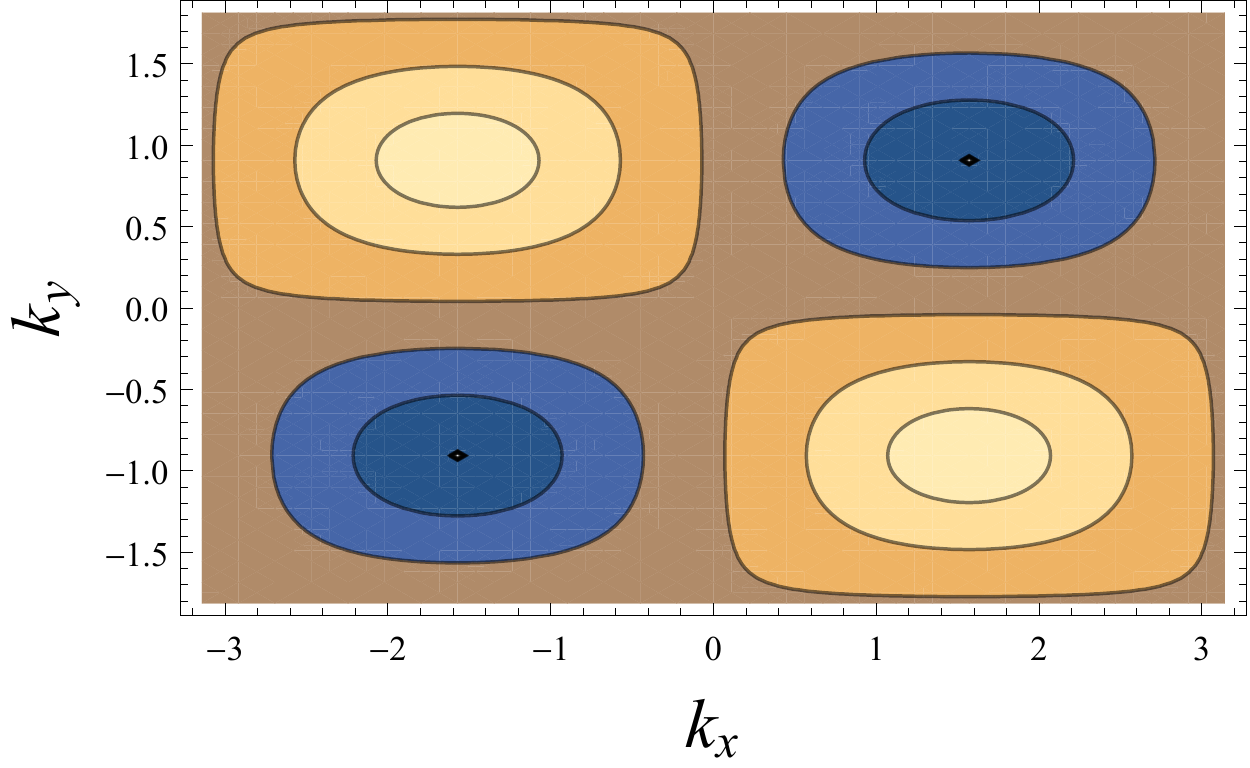}%
}&
\multirow{-3}[2.5]{*}{\subfloat{
  \includegraphics[height=5cm]{Legend}}}\\
\subfloat[$m=2\pi/3$]{%
  \includegraphics[width=.3\linewidth]{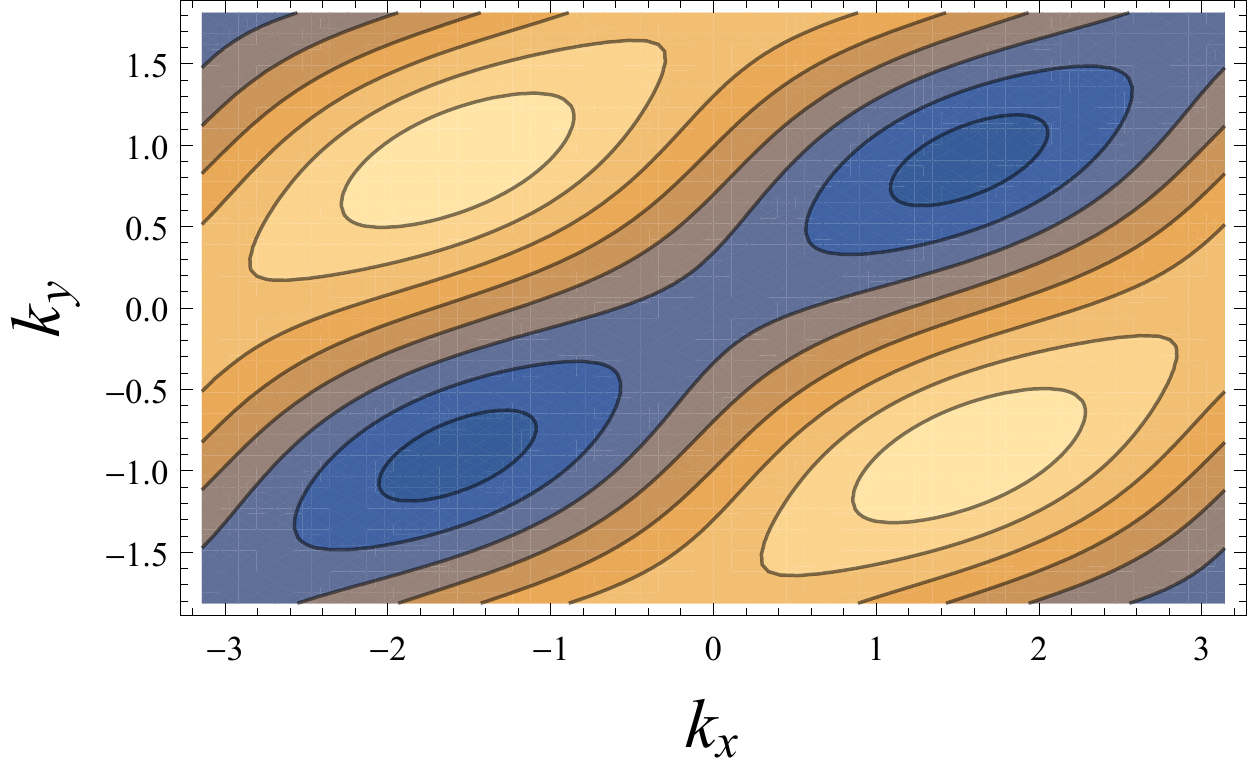}%
}&
\subfloat[$m=\pi$]{%
  \includegraphics[width=.3\linewidth]{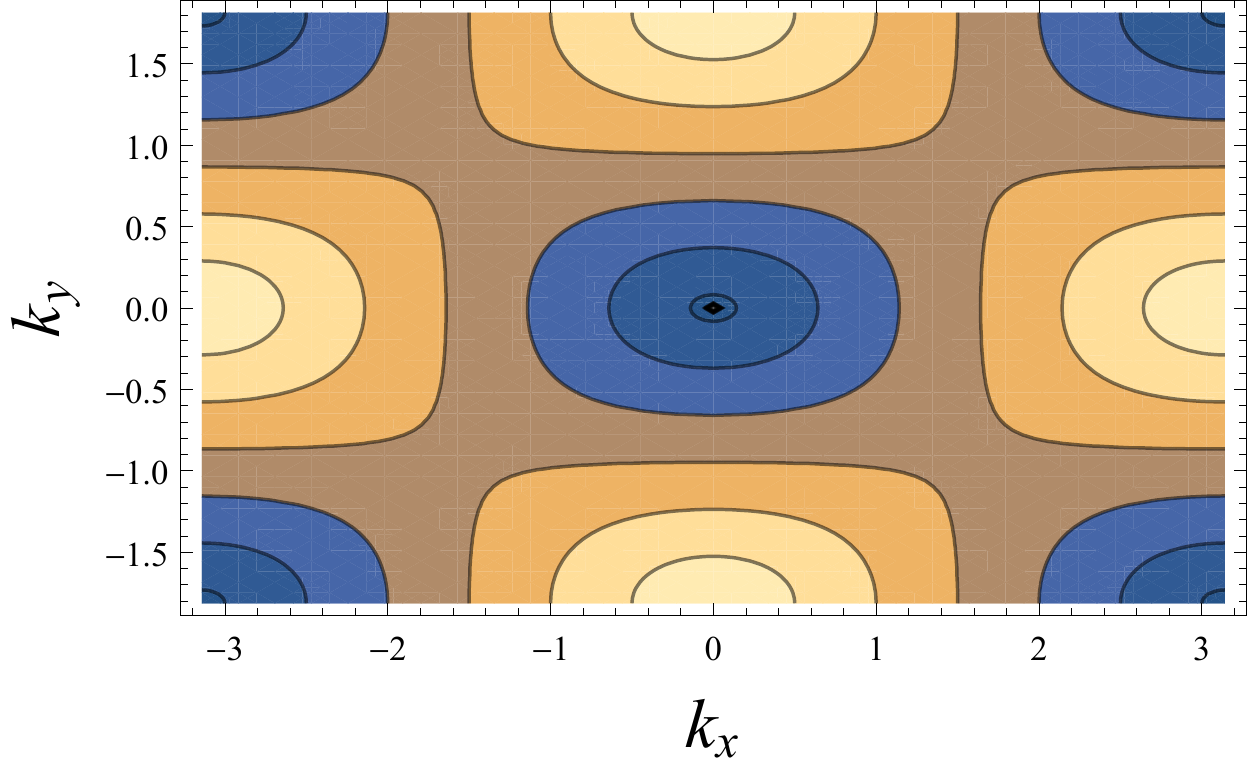}%
}&
\subfloat[$m=4\pi/3$]{%
  \includegraphics[width=.3\linewidth]{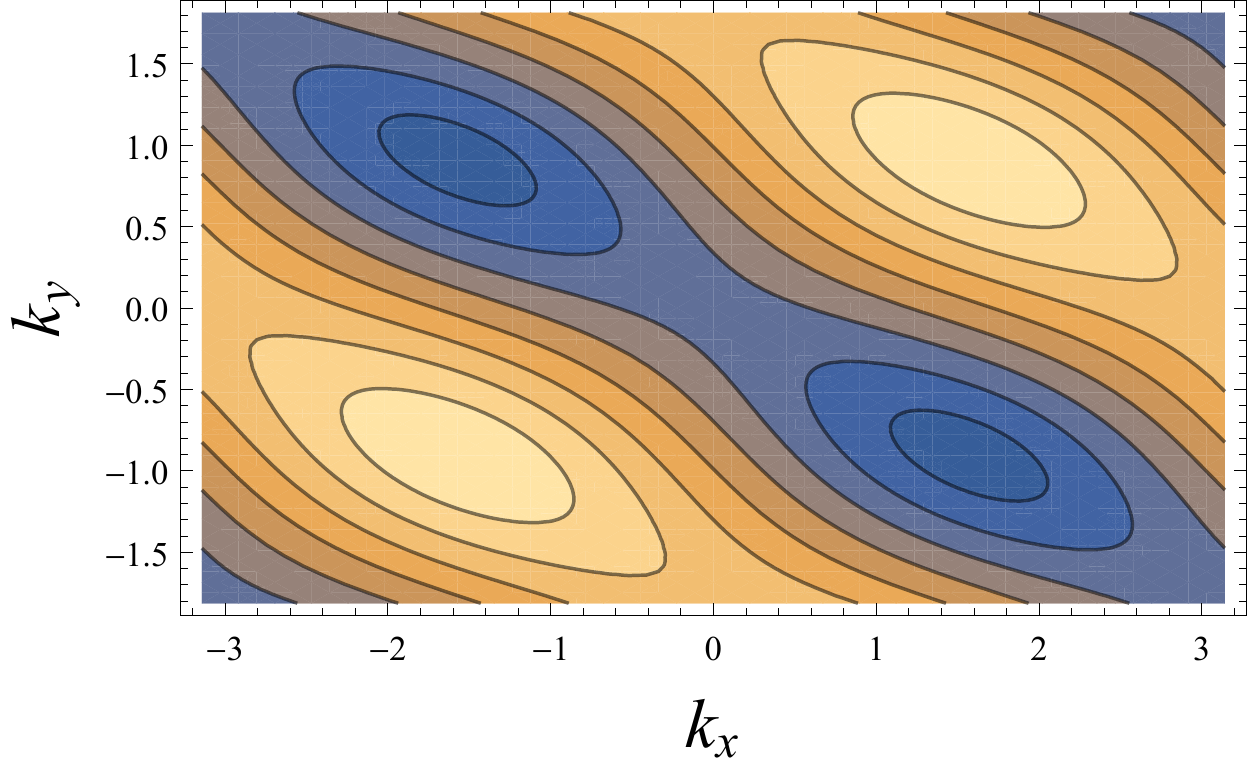}
}\\
\end{tabular}\caption{Dispersion Relations $\omega(\bf{k})$ for the DQW on the isosceles triangular lattice}
\label{IsTri}
\end{figure*}
\begin{figure*}[t]
\begin{tabular}{cccc}
  \subfloat[$m=0$]{%
  \includegraphics[width=.3\linewidth]{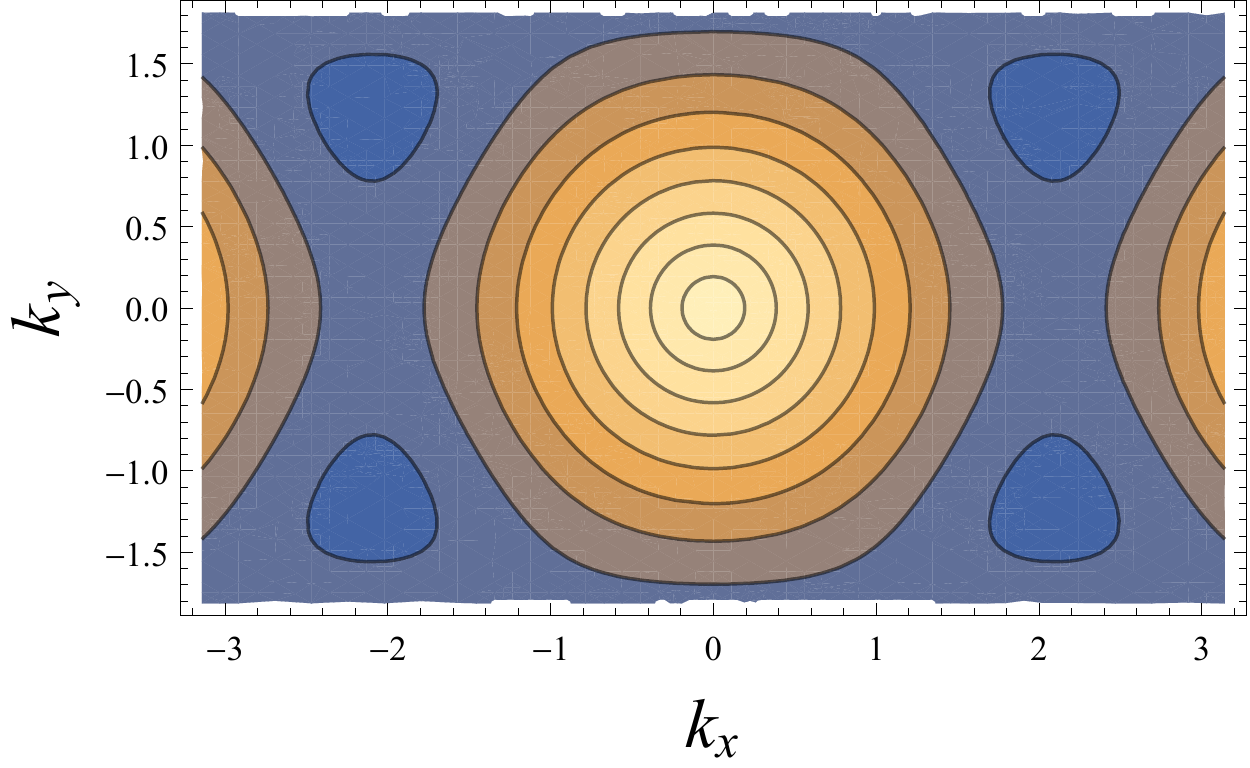}%
}&
\subfloat[$m=2\pi/3\sqrt{3}$]{%
  \includegraphics[width=.3\linewidth]{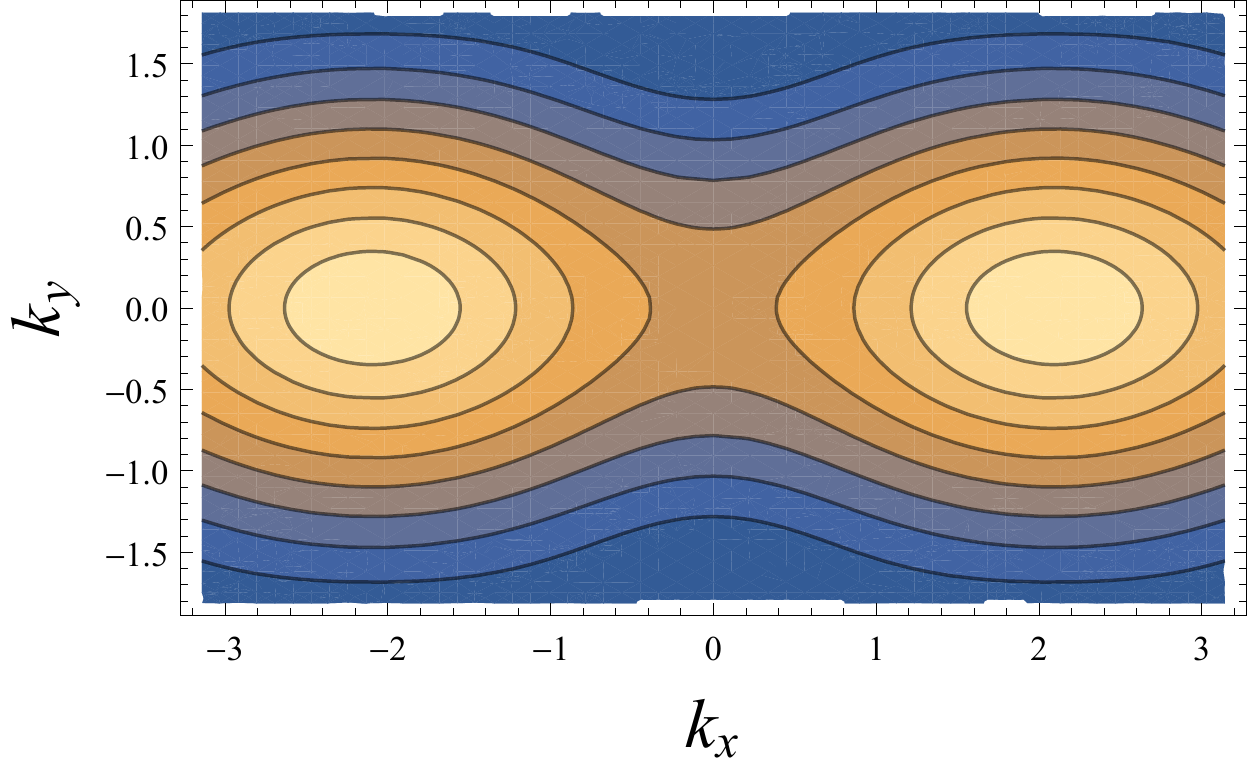}%
}&
\subfloat[$m=\pi/\sqrt{3}$]{%
  \includegraphics[width=.3\linewidth]{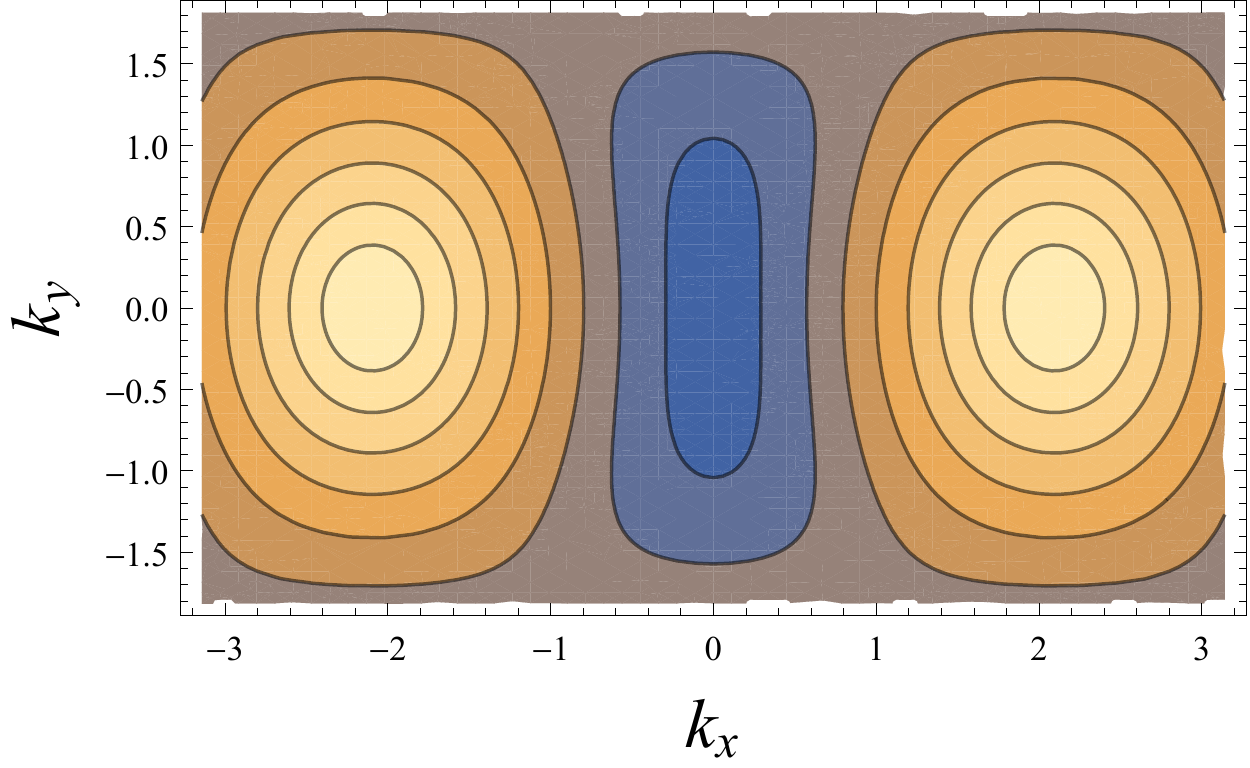}%
}&
\multirow{-3}[2.5]{*}{\subfloat{
  \includegraphics[height=5cm]{Legend}}}\\
\subfloat[$m=4\pi/3\sqrt{3}$]{%
  \includegraphics[width=.3\linewidth]{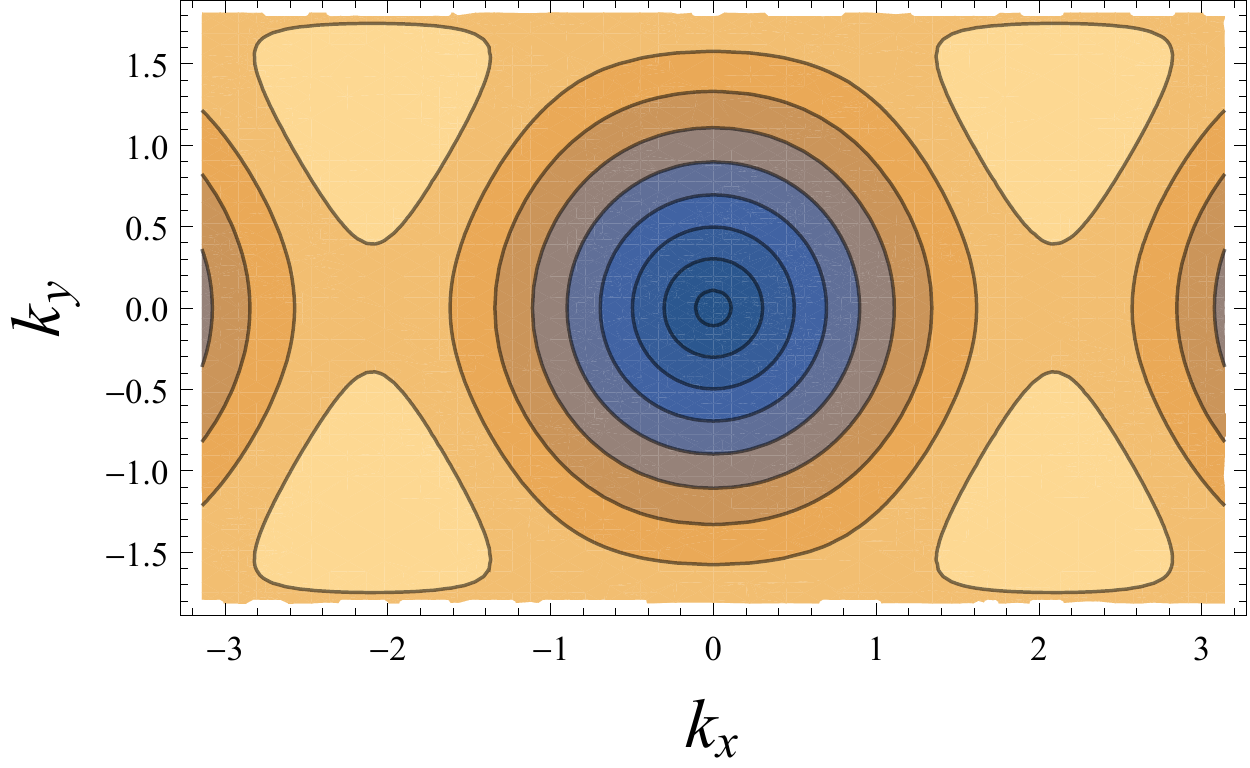}%
}&
\subfloat[$m=2\pi/\sqrt{3}$]{%
  \includegraphics[width=.3\linewidth]{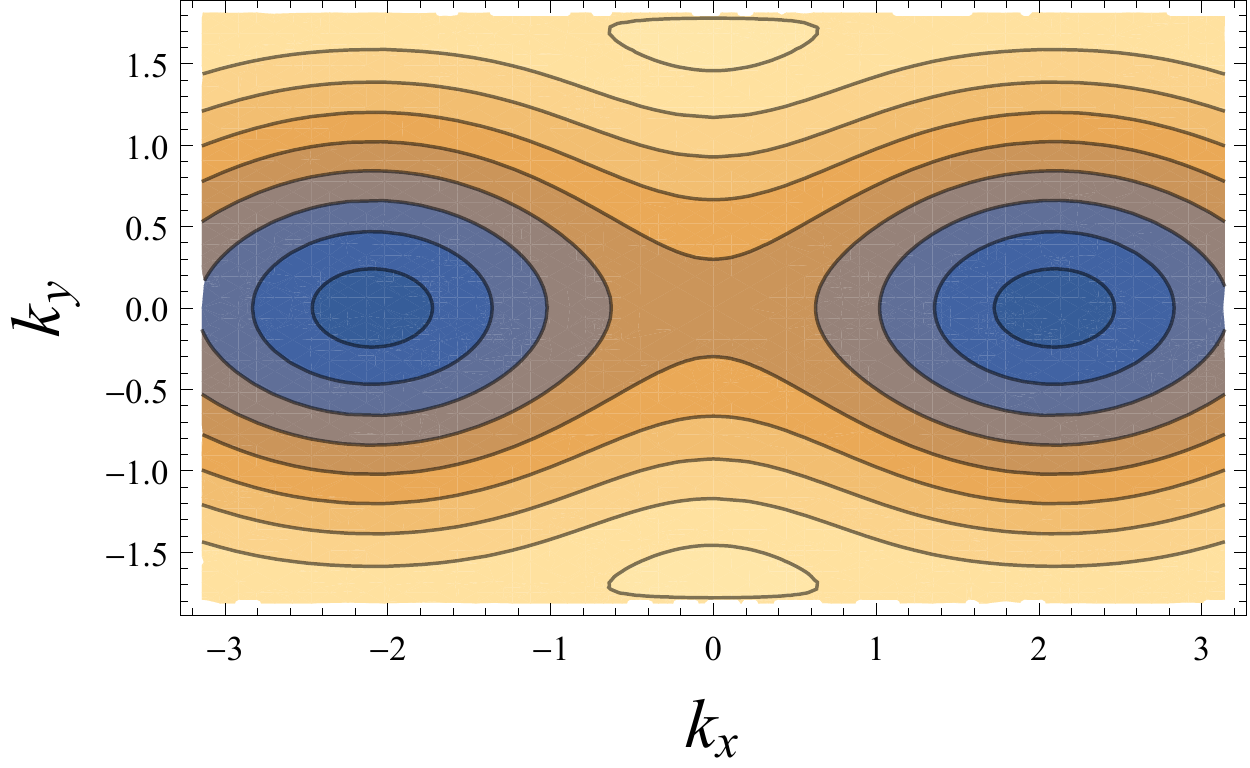}%
}&
\subfloat[$m=8\pi/3\sqrt{3}$]{%
  \includegraphics[width=.3\linewidth]{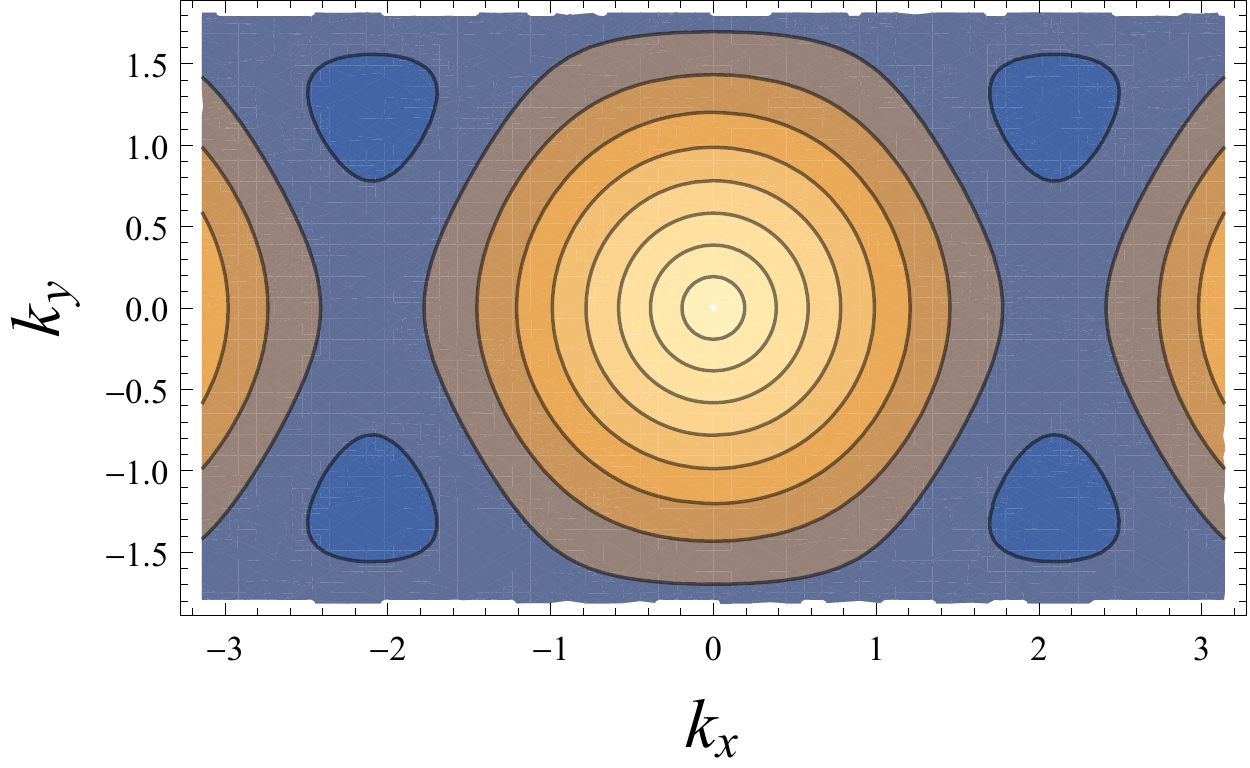}
}\\
\end{tabular}\caption{Dispersion Relations $\omega(\bf{k})$ for the DQW on the hexagonal honeycomb lattice}
\label{HComb}
\end{figure*}

\section{III. Dispersion relations and Zitterbewegung }

Expanding the discrete equations defining the DQWs furnishes an explicit expression for the value taken by the wave function at time $t + \Delta t$ and point $\bf X$ in terms of the values taken by the wave function at time $t$ and various neighbouring points. Searching for linearly polarized plane wave solutions of the form $\Psi(t, {\bf X}) = A(\omega, {\bf k}) \exp \left( i \left(\omega t - {\bf k} \cdot {\bf X} \right) \right)$ delivers a homogeneous linear system for the components of the polarization spinor $ A(\omega, {\bf k})$. This system has a non-vanishing solution only if its determinant vanishes. The coefficients in this system are linear functions of $\Omega =   \exp \left( i \omega \right)$, $K_x =   \exp \left( i k_x \right)$ and $K_y =   \exp \left( i k_y \right)$. The determinant is thus a quadratic function of $\Omega$ and equating this determinant to zero thus furnishes a quadratic equation for $\Omega$, which can be solved exactly, delivering two solutions $\Omega_{\pm}$ as functions of $K_x$ and $K_y$. Because the walks are unitary, both solutions have unit modulus and they are also complex conjugate to each other because the walks are time-reversible. One thus obtains exact expressions
$\pm \omega ({\bf k})$ for the two energy branches of each walk.

The contours of the negative energy branch are plotted in Figs. \ref{6StepEqTri}, \ref{3StepEqTri}, \ref{IsTri} and \ref{HComb} for different values of the mass for each of the four walks considered in this article respectively. For each walk, the Brillouin zone in $\bf k$ space
is $( - \pi, + \pi) \times (- \pi/\sqrt{3}, + \pi/\sqrt{3})$. The dispersion relations  are also periodic in the mass $m$, with period $4 \pi/3$ for the two walks on the equilateral triangle lattice and $2 \pi$ for the walk on the isosceles triangle lattice. The mass periodicity on the honeycomb lattice is $8 \pi/3\sqrt{3}$.

Various conclusions can be drawn from the figures. First, one sees that, for each walk, varying the mass greatly influences the dispersion relation, changing for example the number and positions of the maxima and minima. As for the two walks defined on the equilateral triangle lattice, their symmetries are apparent on the dispersion relations, and both sets of contours therefore look very different (though both walks admit the same continuous limit). In particular, even the contours for vanishing mass are very different. The walk defined on the isosceles triangle lattice, whose continuous limit is also the flat space-time Dirac equation, gives rise to yet another set of contours whose evolution with the mass $m$ does not mirror the evolution obtained for the other two walks.

These contours have a direct consequence about the Zitterbewegung exhibited by the three walks. Let us recall that Zitterbewegung happens because of the interference
of positive and negative energy solutions. For the Dirac equation, the two energy branches are $\omega_\pm = \pm \sqrt { {\bf k}^2 + m^2}$. Zitterbewegung thus happens at frequencies larger than the minimal energy gap $2m$, which is reached for  ${\bf k} = 0$. Figs. \ref{6StepEqTri}, \ref{3StepEqTri} and \ref{IsTri} clearly show that, for non-vanishing values of $m$, the minimal energy gap of three walks considered in this article is inferior to $2m$ and is reached for non-vanishing values of the wave-vector ${\bf k}$. Thus, for non-vanishing mass, Ziterbewegung of frequency lower than $2m$ can be observed on wave-packets centered on non-vanishing values
of ${\bf k}$. For vanishing mass, the six-step walk on the equilateral triangle lattice behaves exactly as solutions of the massless Dirac equation {\sl i.e.} there is no minimal frequency for Zitterbewegung, and the energy gap vanished only at ${\bf k} = 0$. At vanishing mass, the other two walks exhibit a slightly different behaviour. There is still is minimal frequency for Zitterbewegung, but the energy gap vanishes, not only at ${\bf k} = 0$, but also on the four corners of the Brillouin zone. The minimal Zitterbewegung frequencies for the three triangular DQWs for various values of the mass are given in Tables \ref{tab:ZitTabEq3Step}, \ref{tab:ZitTabEq6Step} and \ref{tab:ZitTabIsri}.

\begin{table}
  \centering
  \begin{tabular}{|c|c|c|c|}
    \hline
    $m$&$\omega_m$&$2\omega_m$&$(k_x,k_y)$\\\hline
    \multirow{5}{*}{$0$}&\multirow{5}{*}{$0$}&\multirow{5}{*}{$0$}&$(0,0)$\\
    &&&$(3.14159,1.8138)$\\
    &&&$(3.14159,-1.8138)$\\
    &&&$(-3.14159,1.8138)$\\
    &&&$(-3.14159,-1.8138)$\\\hline
    \multirow{4}{*}{$\frac{\pi}{3}$}&\multirow{4}{*}{$0.451188$}&\multirow{4}{*}{$0.902377$}&$(-2.35619,-0.6046)$\\
    &&&$(2.35619,0.6046)$\\
    &&&$(-0.785398,-1.2092)$\\
    &&&$(0.785398,1.2092)$\\\hline
    \multirow{4}{*}{$\frac{\pi}{2}$}&\multirow{4}{*}{$0.225893$}&\multirow{4}{*}{$0.451789$}&$(-2.34159,-0.383799)$\\
    &&&$(2.34159,0.383799)$\\
    &&&$(-0.831593,-1.4238)$\\
    &&&$(0.831593,1.4238)$\\\hline
    \multirow{6}{*}{$\frac{2\pi}{3}$}&\multirow{6}{*}{$0.812756$}&\multirow{6}{*}{$1.62551$}&$(-2.0944,0)$\\
    &&&$(2.0944,0)$\\
    &&&$(-1.0472,1.8138)$\\
    &&&$(-1.0472,-1.8138)$\\
    &&&$(1.0472,1.8138)$\\
    &&&$(1.0472,-1.8138)$\\\hline
    \multirow{2}{*}{$\pi$}&\multirow{2}{*}{$0.523599$}&\multirow{2}{*}{$1.0472$}&$(-1.5708,0.9069)$\\
    &&&$(1.5708,-0.9069)$\\\hline
    \multirow{5}{*}{$\frac{4\pi}{3}$}&\multirow{5}{*}{$0$}&\multirow{5}{*}{$0$}&$(0,0)$\\
    &&&$(3.14159,1.8138)$\\
    &&&$(3.14159,-1.8138)$\\
    &&&$(-3.14159,1.8138)$\\
    &&&$(-3.14159,-1.8138)$\\\hline
  \end{tabular}
  \caption{Minimal frequencies for the equilateral triangle three-step DQW} \label{tab:ZitTabEq3Step}
\end{table}

\begin{table}
  \centering
  \begin{tabular}{|c|c|c|c|}
    \hline
    $m$&$\omega_m$&$2\omega_m$&$(k_x,k_y)$\\\hline
    $0$&$0$&$0$&$(0,0)$\\\hline
    \multirow{2}{*}{$\frac{\pi}{3}$}&\multirow{2}{*}{$1.10603$}&\multirow{2}{*}{$2.21205$}&$(1.37445,0)$\\
    &&&$(-1.37445,0))$\\\hline
    \multirow{2}{*}{$\frac{\pi}{2}$}&\multirow{2}{*}{$0.565516$}&\multirow{2}{*}{$1.13103$}&$(1.8326,0)$\\
    &&&$(-1.8326,0))$\\\hline
    \multirow{4}{*}{$\frac{2\pi}{3}$}&\multirow{4}{*}{$0.505361$}&\multirow{4}{*}{$1.0172$}&$(-1.23095,1.8138)$\\
    &&&$(-1.23095,-1.8138)$\\
    &&&$(1.23095,1.8138)$\\
    &&&$(1.23095,-1.8138)$\\\hline
    \multirow{2}{*}{$\pi$}&\multirow{2}{*}{$0.523599$}&\multirow{2}{*}{$1.0472$}&$(3.14159,k_y)$\\
    &&&$(-3.14159,k_y)$\\\hline
   $\frac{4\pi}{3}$&$0$&$0$&$(0,0)$\\\hline
  \end{tabular}
  \caption{Minimal frequencies for the equilateral triangle six-step DQW} \label{tab:ZitTabEq6Step}
\end{table}

\begin{table}
  \centering
  \begin{tabular}{|c|c|c|c|}
    \hline
    $m$&$\omega_m$&$2\omega_m$&$(k_x,k_y)$\\\hline
    \multirow{5}{*}{$0$}&\multirow{5}{*}{$0$}&\multirow{5}{*}{$0$}&$(0,0)$\\
    &&&$(3.14159,1.8138)$\\
    &&&$(3.14159,-1.8138)$\\
    &&&$(-3.14159,1.8138)$\\
    &&&$(-3.14159,-1.8138)$\\\hline
    \multirow{2}{*}{$\frac{\pi}{3}$}&\multirow{2}{*}{$0.523599$}&\multirow{2}{*}{$1.0472$}&$(-1.5708,0.9069)$\\
    &&&$(1.5708,-0.9069)$\\\hline
    \multirow{2}{*}{$\frac{\pi}{2}$}&\multirow{2}{*}{$0$}&\multirow{2}{*}{$0$}&$(-1.5708,0.9069)$\\
    &&&$(1.5708,-0.9069)$\\\hline
    \multirow{2}{*}{$\frac{2\pi}{3}$}&\multirow{2}{*}{$0.523599$}&\multirow{2}{*}{$1.0472$}&$(-1.5708,0.9069)$\\
    &&&$(1.5708,-0.9069)$\\\hline
    \multirow{4}{*}{$\pi$}&\multirow{4}{*}{$0$}&\multirow{4}{*}{$0$}&$(-3.14159,0)$\\
    &&&$(3.14159,0)$\\
    &&&$(0,-1.8138)$\\
    &&&$(0,1.8138)$\\\hline
    \multirow{2}{*}{$\frac{4\pi}{3}$}&\multirow{2}{*}{$0.523599$}&\multirow{2}{*}{$1.0472$}&$(-1.5708,-0.9069)$\\
    &&&$(1.5708,0.9069)$\\\hline
  \end{tabular}
  \caption{Minimal frequencies for the isosceles triangle DQW} \label{tab:ZitTabIsri}
\end{table}

\section{IV. Gauge invariant coupling to electromagnetic fields}

\subsection{1. Introducing the fields}

The above DQWs are invariant by a global change of phase of the spinor $\Psi$, but they are not invariant under a local change of phase. However, they can all be transformed into locally $U(1)$ gauge invariant. We will now present in detail the principle behind this generalization on the three-step DQW defined on the equilateral lattice. Adding an electromagnetic field to the other three walks can be done in a similar way.

To achieve local $U(1)$ gauge invariance, we define each $U_i$ operator entering the definition of the walk by
%a more general operator ${\mathcal U}_i$ defined by
\begin{equation}
 \left(U_i \Psi \right)(t, {\bf X}) = U \left( \alpha_i(t, {\bf X}),  \xi_i(t, {\bf X}), \zeta_i(t, {\bf X}), 0\right) \Psi(t, {\bf X}),
\end{equation}
where the $\alpha_i$'s, $\xi_i$'s and $\zeta_i$'s are arbitrary functions of $t$ and $\bf X$. Note that the choice $\theta_i = 0$
make the walk actually independent of $\zeta_i$.

Let us break down the definition of the walk in three time sub-steps and write
\begin{eqnarray}
\Psi(t + \Delta t/3) & = & U_1(t) {\tilde S}_1 \Psi (t), \nonumber \\
\Psi(t +  2\Delta t/3) & = & {\tilde R}_1^{-1} U_2(t) {\tilde S}_2 {\tilde R}_1 \Psi (t + \Delta t/3), \nonumber \\
\Psi(t + \Delta t) & = & {\tilde R}_2^{-1} U_3(t)  {\tilde S}_3 {\tilde R}_2 \Psi (t + 2\Delta t/3).
\label{eq:DTQW3steps}
\end{eqnarray}
Let $\phi$ be the phase of the spinor $\Psi$ and consider an arbitrary local change of phase
$\phi'(t + r \Delta t/3, {\bf X}) = \phi(t + r \Delta t/3, {\bf X}) + \delta  \phi(t + r \Delta t/3, {\bf X})$ for all
$(t, {\bf X})$ and $r = 0, 1, 2$. Note that one thus allows for an arbitrary change of phase, not
only at all integer time steps, but also at all intermediate time sub-steps.
We accompany this arbitrary change of phase by a change in the functions $\alpha_i$, $\xi_i$ and $\zeta_i$, introducing $\alpha'_i (t, {\bf X}) = \alpha_i (t, {\bf X}) + \delta \alpha_i (t, {\bf X})$ etc.

Focus now on the first intermediate evolution equation between time $t$ and time
$t + \Delta t/3$. A direct computation identical to the one already presented for DQWs
in $2D$ space-times reveals that this equation is identical for the primed and the unprimed walk provided
\begin{eqnarray}
\delta \alpha_1 (t, {\bf X})  & =  & \delta \phi(t+ \Delta t/3, {\bf X})  - \sigma_1(t, {\bf X}),  \nonumber \\
\delta \xi_1 (t, {\bf X}) & = & - \delta_1 (t, {\bf X}),% \nonumber \\
%\delta \zeta_1 (t, {\bf X}) & =  & + \delta_1 (t, {\bf X})
\end{eqnarray}
where
$\sigma_1(t, {\bf X}) = (\delta \phi(t, {\bf X}_1) + \delta \phi(t, {\bf X}_{4}))/2$ and $\delta_1 (t, {\bf X}) = (\delta  \phi(t, {\bf X}_1) - \delta  \phi(t, {\bf X}_{4}))/2$. There is no constraint on $\zeta_i'$
because it actually does not enter the definition of the walk.

Let us now proceed to the second time sub-step. It seems like the situation is more complex but it is
actually identical, because a local change of phase affects both components of the spinor identically and, thus, commute with the operator $R_1$. One thus obtains local gauge invariance at this time sub-step if
\begin{eqnarray}
\delta \alpha_2 (t, {\bf X})  & =  & \delta \phi(t+ 2 \Delta t/3, {\bf X})  - \sigma_2(t, {\bf X}),  \nonumber \\
\delta \xi_2 (t, {\bf X}) & = & - \delta_2 (t, {\bf X}),% \nonumber \\
%\delta \zeta_2 (t, {\bf X}) & =  & + \delta_2 (t, {\bf X})
\end{eqnarray}
where
$\sigma_2(t, {\bf X}) = (\delta \phi(t + \Delta t/3, {\bf X}_2) + \delta  \phi(t + \Delta t/3, {\bf X}_{5}))/2$ and $\delta_2 (t, {\bf X}) = (\delta  \phi(t + \Delta t/3, {\bf X}_2) - \delta  \phi(t + \Delta t/3, {\bf X}_{5}))/2$.
The same reasoning goes for the third and final time sub-step and gauge invariance at this sub-step is provided if
\begin{eqnarray}
\delta \alpha_3 (t, {\bf X})  & =  & \delta \phi(t+ \Delta t, {\bf X}) - \sigma_3(t, {\bf X}),  \nonumber \\
\delta \xi_3 (t, {\bf X}) & = & - \delta_3 (t, {\bf X}),% \nonumber \\
%\delta \zeta_3 (t, {\bf X}) & =  & + \delta_3 (t, {\bf X})
\end{eqnarray}
where $\sigma_3(t, {\bf X}) = (\delta  \phi(t + 2 \Delta t/3, {\bf X}_3) + \delta  \phi(t + 2 \Delta t/3, {\bf X}_{6}))/2$ and $\delta_3 (t, {\bf X}) = (\delta \phi (t + 2 \Delta t/3, {\bf X}_3) - \delta  \phi(t + 2\Delta t/3, {\bf X}_{6}))/2$.

One way to acknowledge that sub-steps are just sub-steps, and not full time-steps, is to consider phase changes such that $\delta \phi(t, {\bf X}) = \delta \phi(t+ \Delta t/3, {\bf X}) =  \delta \phi(t+2 \Delta t/3, {\bf X})$ (compare in particular with the definition of the $\alpha_i$'s, $\xi_i$'s and $\zeta_i$'s, which are indexed by $t$, though all describe what happens between intermediate sub-steps). For these phase changes, the above equations simplify into
\begin{eqnarray}
\delta \alpha_3 (t, {\bf X})  & =  & \delta \phi(t+ \Delta t, {\bf X}) - \sigma_3(t, {\bf X}),  \nonumber \\
\delta \alpha_2 (t, {\bf X})  & =  & \delta \phi(t, {\bf X}) - \sigma_2(t, {\bf X}),  \nonumber \\
\delta \alpha_1 (t, {\bf X})  & =  & \delta \phi(t, {\bf X}) - \sigma_1(t, {\bf X}),  \nonumber \\
\delta \xi_i (t, {\bf X}) & = & - \delta_i (t, {\bf X}),% \nonumber \\
%\delta \zeta_i (t, {\bf X}) & =  & + \delta_i (t, {\bf X})
\end{eqnarray}
where
$\sigma_i(t, {\bf X}) = (\delta  \phi(t, {\bf X}_i) + \delta \phi(t, {\bf X}_{i + 3}))/2$, $\delta_i (t, {\bf X}) = (\delta \phi(t, {\bf X}_i) - \delta \phi(t, {\bf X}_{i + 3}))/2$ for $i = 1, 2, 3$.

Another natural classes of phase changes is defined by the relations
\begin{eqnarray}
\delta \phi(t+ \Delta t/3, {\bf X}) &-& \delta \phi(t, {\bf X}) \nonumber \\
& = & \frac{1}{3} \left( \delta \phi(t+ \Delta t, {\bf X}) - \delta \phi(t, {\bf X})\right), \nonumber \\
\delta \phi(t+ 2\Delta t/3, {\bf X}) &-& \delta \phi(t + \Delta t/3, {\bf X}) \nonumber \\
& = & \frac{1}{3} \left( \delta \phi(t+ \Delta t, {\bf X}) - \delta \phi(t, {\bf X})\right).\nonumber\\
%\delta \alpha_3 (t, {\bf X})  & =  & \delta \phi(t, {\bf X}) - \delta \phi(t+ \Delta t, {\bf X}) + \sigma_3(t, {\bf X})  \nonumber \\
%\delta \alpha_2 (t, {\bf X})  & =  & + \sigma_2(t, {\bf X})  \nonumber \\
%\delta \alpha_1 (t, {\bf X})  & =  & + \sigma_1(t, {\bf X})  \nonumber \\
%\delta \xi_i (t, {\bf X}) & = & - \delta_i (t, {\bf X}) \nonumber \\
%\delta \zeta_i (t, {\bf X}) & =  & + \delta_i (t, {\bf X})
\end{eqnarray}
For these phase changes, the gauge transformation of the $\alpha_i$'s and $\xi_i$'s simplifies into
\begin{eqnarray}
\delta \alpha_i (t, {\bf X})  & =  & - \frac{1}{3} \left( \delta \phi(t, {\bf X})\right) - \sigma_i(t, {\bf X}),  \nonumber \\
\delta \xi_i (t, {\bf X}) & = & - \delta_i (t, {\bf X}),% \nonumber \\
%\delta \zeta_i (t, {\bf X}) & =  & + \delta_i (t, {\bf X})
\end{eqnarray}
for $i = 1, 2, 3$.
 
To perform the continuous limit, we write $\alpha_i = \epsilon {\bar \alpha}_i$, $\xi_i = \epsilon {\bar \xi}_i$, $\zeta_i = \epsilon {\bar \zeta}_i$ and let $\epsilon$ tend to zero, keeping all bar angles finite.
The DQW dynamics then delivers the Dirac equation with electromagnetic potential
\begin{eqnarray}
A_0 & = & \frac{2}{3} ({\bar \alpha}_1+{\bar \alpha}_2+ {\bar \alpha}_3),  \nonumber \\
A_1 & =- & \frac{2}{3} \left({\bar \xi}_1+ \frac{1}{2} \left({\bar \xi}_2- {\bar \xi}_3 \right) \right),   \nonumber \\
A_2 & =- & \frac{1}{\sqrt{3}} \left({\bar \xi}_2 + {\bar \xi}_3  \right).
\end{eqnarray}

\begin{figure*}[t]
\begin{tabular}{ccc}
  \subfloat[$E_x=0$]{%
  \includegraphics[width=.3\linewidth]{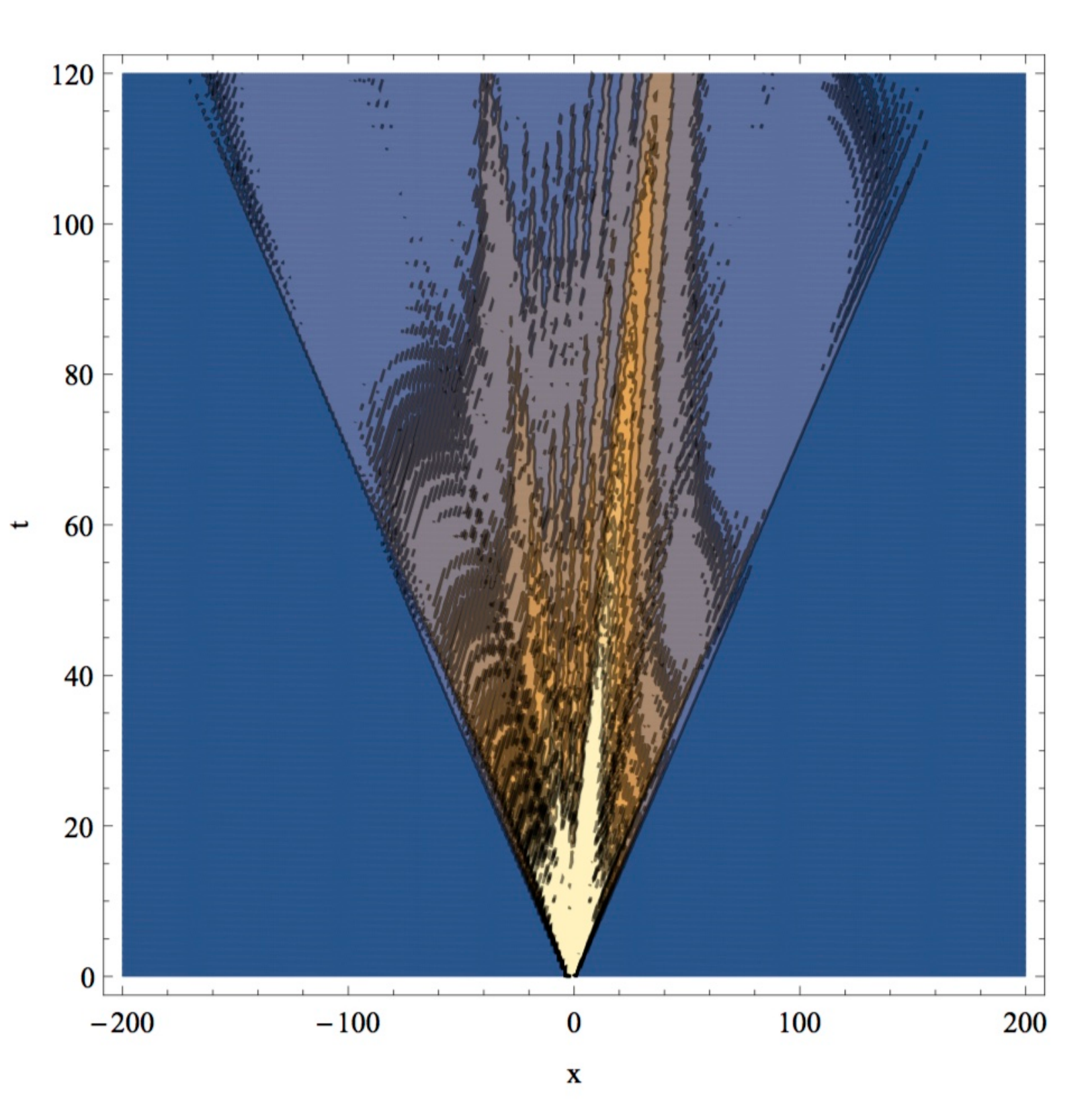}%
\includegraphics[width=1cm]{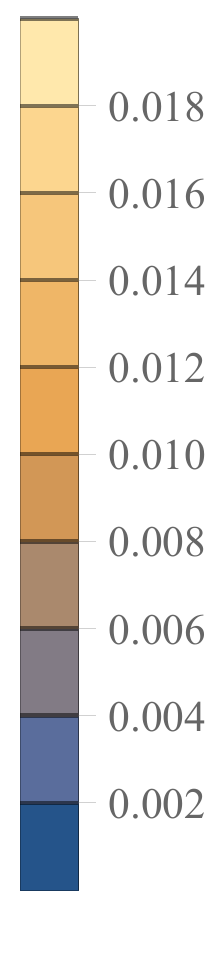}%
}&\subfloat[$E_x=0.1$]{%
  \includegraphics[width=.3\linewidth]{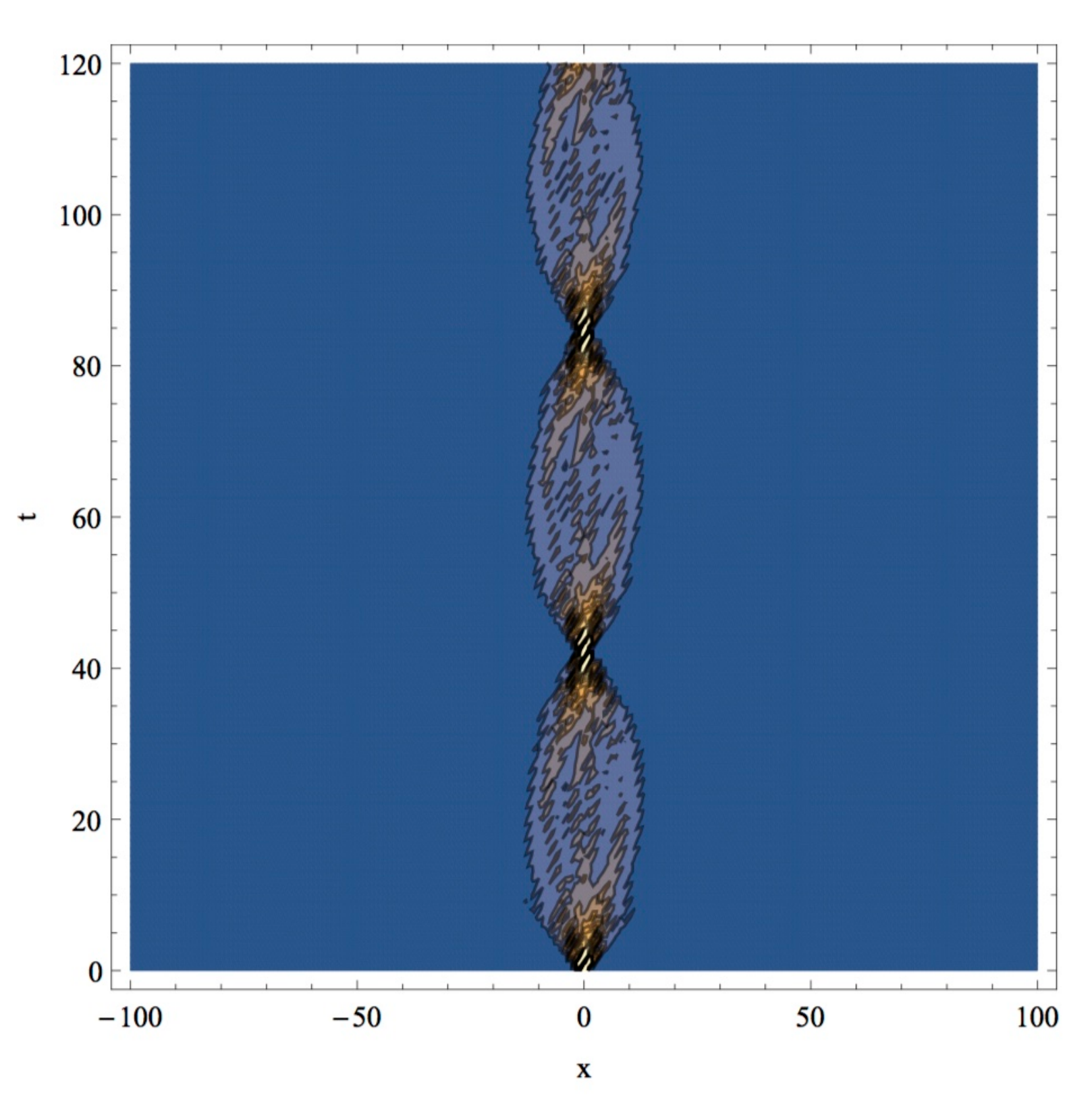}%
\includegraphics[width=1cm]{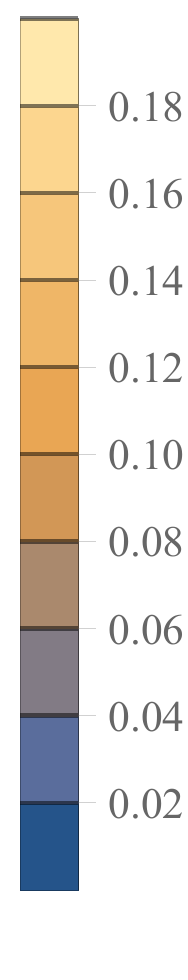}%
}&
\subfloat[$E_x=0.05$]{%
  \includegraphics[width=.3\linewidth]{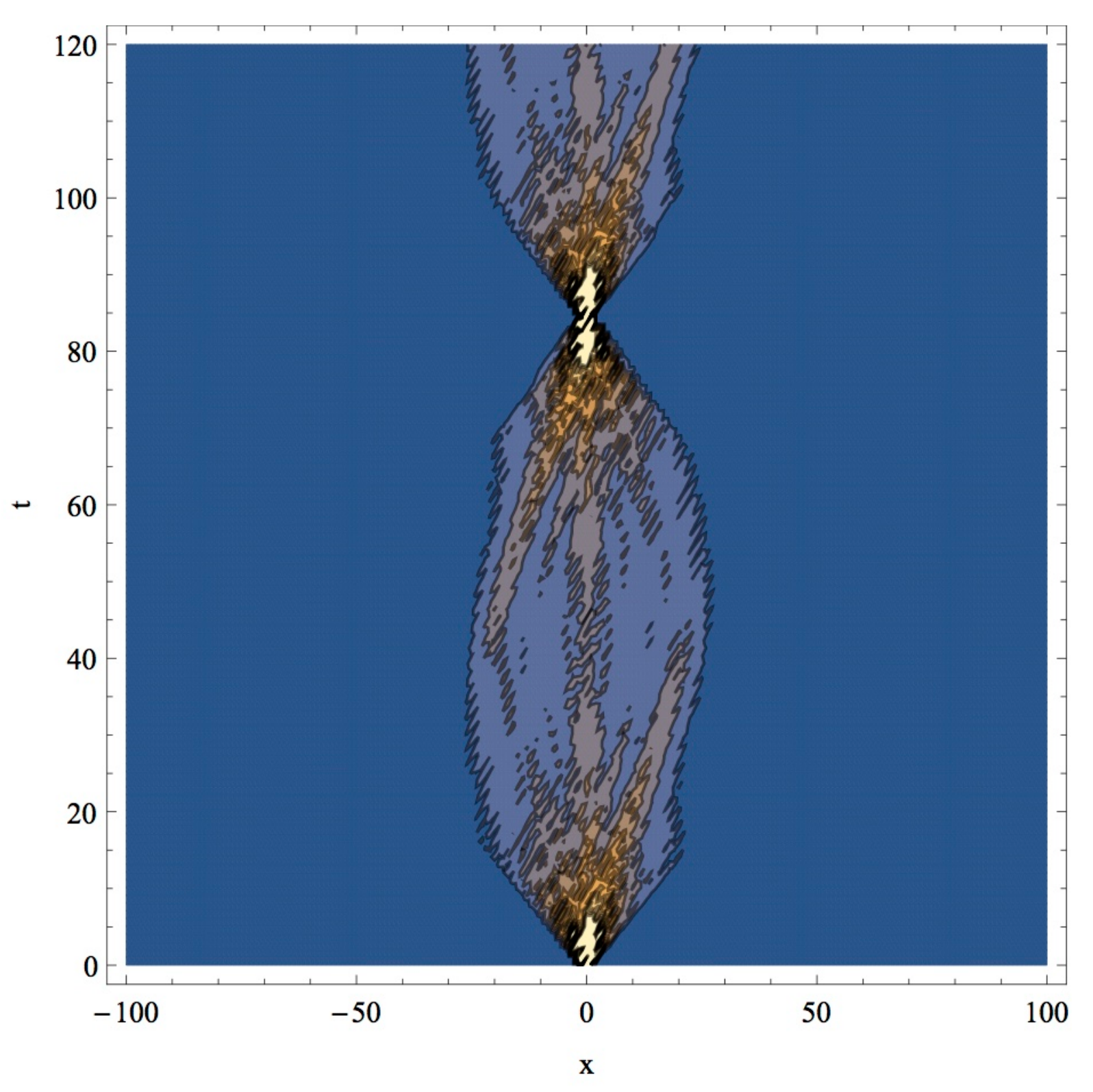}%
\includegraphics[width=1cm]{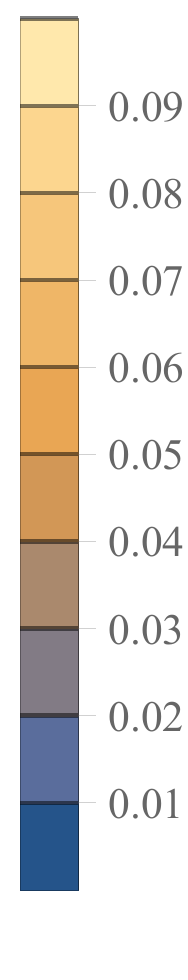}%
}\\
\end{tabular}\caption{Density plots of (a) no electric field (free space), and (b-c) electric field in the $x$-direction}
\label{DensEx}
\end{figure*}

\begin{figure}[t]
\includegraphics[width=.65\linewidth]{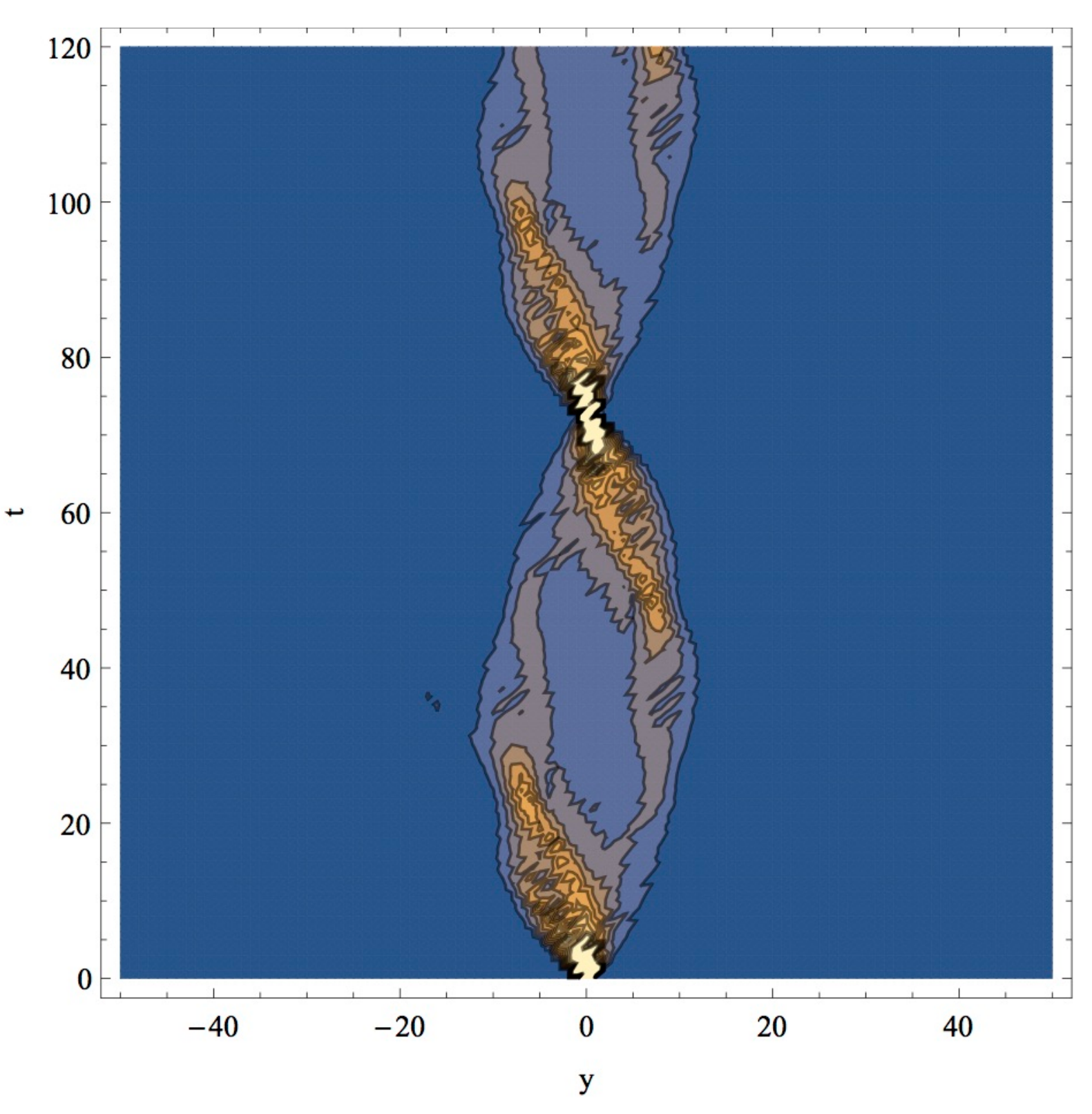}%
\includegraphics[width=1cm]{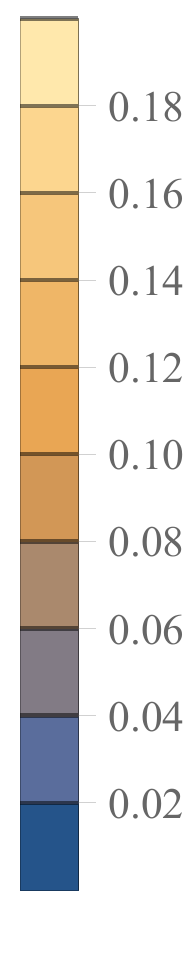}%
\caption{Density plot of electric field in the $y$-direction: $E_y=0.1$}
\label{DensEy}
\end{figure}

\subsection{2. Motion in a uniform constant electric field}

Based on the above relations between $(A_0, A_1, A_2)$ and the phase variables $\alpha_i$ and $\xi_i$ a simple way to couple the DQW to a constant homogeneous electric field of cartesian components $E_1$ and $E_2$ is to choose identically vanishing $\alpha_i$'s and to impose $\xi_1 = 0$ together with
\begin{eqnarray}
\xi_2 & = & \frac{1}{2} \, \left( - 3 E_x + \sqrt{3} E_y \right), \nonumber \\
\xi_3 & = & \frac{1}{2} \, \left( + 3 E_x + \sqrt{3} E_y \right).
\end{eqnarray}
Consider first the case $E_y = 0$. Figure \ref{DensEx} presents contour plots of the walk projected unto the $(t, x)$ for various values of $E_x$ and for the symmetric initial condition $\Psi = (1/\sqrt{2}, 1/\sqrt{2})^\top$. Note that this initial condition probes the walk dynamics outside the continuous limit.

In the presence of a non-vanishing electric field, the spreading characteristic of the free walk is replaced by apparent oscillations. These can be understood qualitatively by recalling that, in one spatial dimension, a quantum particle moving in continuous space-time submitted to the action of a uniform homogeneous electric field $E$ combined with a periodic potential undergo oscillations \cite{hartmann2004dynamics}. The period $T_B$ of these oscillations is called the Bloch period and $T_B = 2 \pi/E$ as the Brillouin zone is of length $2 \pi$. Without getting into detailed computations, the periodicity of the potential generates a finite-sized Brillouin zone in momentum space and the period of the oscillations correspond to the time necessary for the particle to cross the Brillouin zone. In the case presented in Figure \ref{DensEx}, there is evidently no periodic potential, but the quantum walk admits a finite-sized Brillouin zone of length $2 \pi$ along the $x$-axis. The apparent period coincides exactly with the Bloch period, if one takes into account the $3/2$ proportionality factor between the time $t$ used to parametrise the walk and the standard choice of continuous time in the Dirac equation (see above).

\begin{figure*}[tph]
\begin{tabular}{cc}
  \subfloat[]{%
  \includegraphics[width=.45\linewidth]{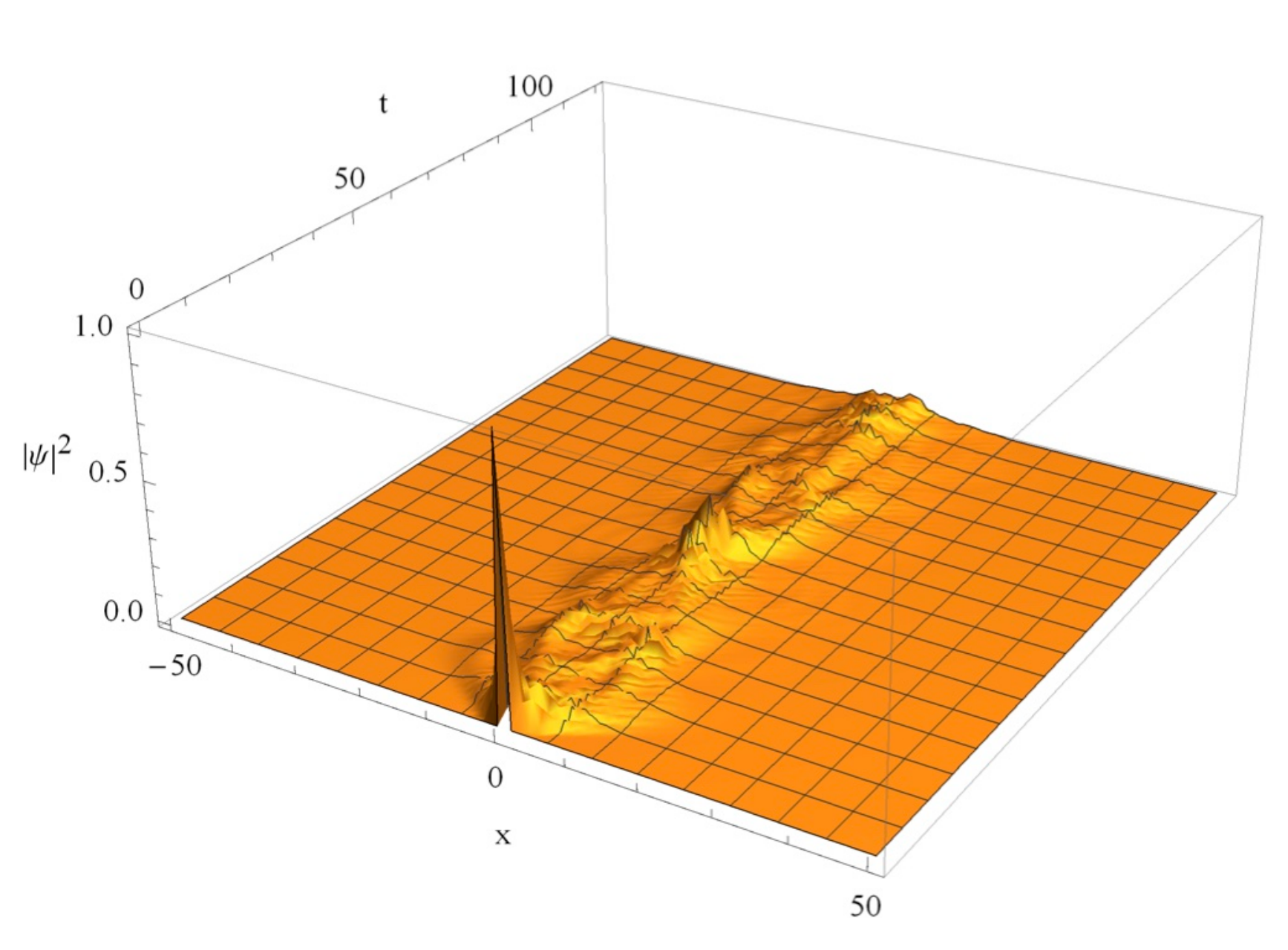}%
}&\subfloat[]{%
  \includegraphics[width=.48\linewidth]{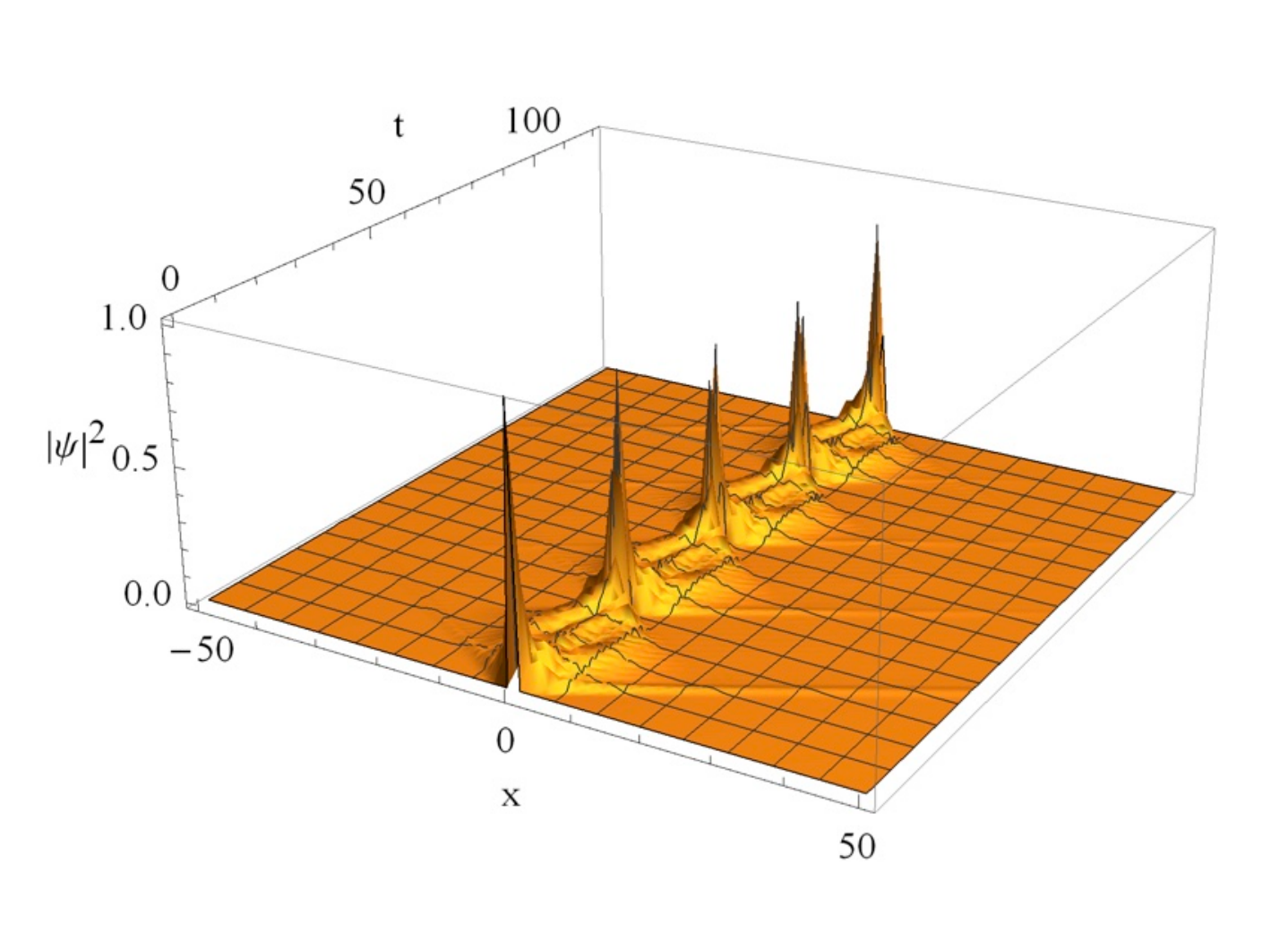}%
}\\
\end{tabular}\caption{Density plots of (a) electric field in both $x$-direction and $y$-direction: $E_x=0.1$ and $E_y=0.1$ and (b) electric field only in $x$-direction of equivalent magnitude ($E_x=\frac{\sqrt{2}}{10}$)}
\label{DensMixed}
\end{figure*}

This interpretation of the observed oscillations can be confirmed by two numerical experiments. First choose now an electric field in the $y$ direction. The length of the Brillouin zone in the $y$-direction
is $2 \pi/\sqrt{3}$, so the Bloch period is now $T_B^y = T_B/\sqrt{3}$. Oscillations at this period are indeed observed on Figure \ref{DensEy}. Choose now an electric field with equal components in the $x$ and $y$ direction. Since the lengths of the $x$- and $y$- the Brillouin zones differ by a multiplicative factor $\sqrt{3}$, their ratio is not a rational number and there is no reason to expect any periodicity in the walk. This is indeed displayed in Figure \ref{DensMixed}, which is obtained for ${\bf E} = (1, 1)$.

\section{V. Conclusion}

We have proposed four different DQWs on two regular triangular lattices and one hexagonal honeycomb lattice, and proved that they all admit the free Dirac equation as continuous limits. We have analyzed in depth the qualitative difference between the four walks, and their consequences for Zitterbewegung. We have also shown that these walks can be extended to incorporate a gauge-invariant coupling to discrete electromagnetic fields and addressed Bloch oscillations. These results show that DQWs defined on more general lattices than the square lattice can be used to study the free Dirac dynamics and that coupling the walks to arbitrary electromagnetic fields in a gauge invariant manner is also possible.

Many interesting questions still remain open. The first class of extensions addresses questions on triangular lattices. For example, can one incorporate other Yang-Mills fields and gravity on triangular lattices? And can one define gauge invariant field strengths for the Yang-Mills fields and gravity? Also, what about irregular triangular lattices, for example with defects?
The second set of questions addresses other lattices and more general discrete structures. Can the results presented in this work be extended to arbitrary regular and non-regular lattices on the $2D$ plane and in spaces of higher dimensions? Do they also translate on graphs, at least regular ones? And can one couple DQWs on graphs with Yang-Mills fields and gravity? Such questions will be addressed in our future work. 

\noindent
{\sl Note added}: Another research team from France and Spain has worked independently of us on DQWs on triangle and honeycomb lattices and 
we exchanged our manuscripts after both were completed. The other team's manuscript offers a very clear and constructive 
presentation of two quantum walks which converge to the Dirac equation as the space-time step goes to zero. These two walks are particularly elegant 
because they are built out of identical sub-steps. We feel both manuscripts complement each other nicely and therefore recommend the other manuscript 
to the reader's attention. 

\section{Acknowledgments}

%Fabrice: please add any acknowledgments you may have before the following.
GJ and JBW would like to thank Ian McArthur for useful discussions on quantum walks and Dirac dynamics in general.

\bibliographystyle{apsrev4-1}
\bibliography{Biblio.bib}
\end{document}